\documentclass[aps, pra, reprint, groupedaddress, floatfix]{revtex4-1} 

\usepackage{graphicx}
\usepackage{epstopdf}
\usepackage{bm, amsmath, amssymb}
\usepackage{float}
\usepackage{placeins}
\usepackage{xcolor}
\usepackage{siunitx}
\usepackage[T1]{fontenc}

\newcommand{\figwidth}{3.375in} 
\newcommand{\blue}[1]{\textcolor{black}{#1}}

\begin{document}

\title{Optimizing the nonlinearity and dissipation of a SNAIL Parametric Amplifier for dynamic range}

\author{N. E. Frattini}
\thanks{Equal contribution}
\email{nicholas.frattini@yale.edu}
\affiliation{Department of Applied Physics, Yale University, New Haven, CT 06520, USA}
\author{V. V. Sivak}
\thanks{Equal contribution}
\email{vladimir.sivak@yale.edu}
\affiliation{Department of Applied Physics, Yale University, New Haven, CT 06520, USA}
\author{A. Lingenfelter}
\affiliation{Department of Applied Physics, Yale University, New Haven, CT 06520, USA}
\author{S. Shankar}
\affiliation{Department of Applied Physics, Yale University, New Haven, CT 06520, USA}
\author{M. H. Devoret}
\email{michel.devoret@yale.edu}
\affiliation{Department of Applied Physics, Yale University, New Haven, CT 06520, USA}

\date{\today}

\begin{abstract}
We present a quantum-limited Josephson-junction-based 3-wave-mixing parametric amplifier, the SNAIL Parametric Amplifier (SPA), which uses an array of SNAILs (Superconducting Nonlinear Asymmetric Inductive eLements) as the source of tunable nonlinearity. We show how to engineer the nonlinearity over multiple orders of magnitude by varying the physical design of the device.
As a function of design parameters, we systematically explore two important amplifier nonidealities that limit dynamic range: the phenomena of gain compression and intermodulation distortion, whose minimization are crucial for high-fidelity multi-qubit readout.
Through a comparison with first-principles theory across multiple devices, we demonstrate how to optimize both the nonlinearity and the input-output port coupling of these SNAIL-based parametric amplifiers to achieve higher saturation power, without sacrificing any other desirable characteristics.
The method elaborated in our work can be extended to improve all forms of parametrically induced mixing that can be employed for quantum information applications.
\end{abstract}

\maketitle

\section{Introduction}\label{sec:Intro}
Quantum-limited Josephson parametric amplifiers \cite{vijay_invited_2009,roy_introduction_2016} are a key component in many precision microwave measurement setups such as for the readout of superconducting qubits \cite{johnson_heralded_2012,riste_initialization_2012,hatridge_quantum_2013, jeffrey_fast_2014,walter_rapid_2017}, the high-sensitivity detection of electron spin resonance \cite{bienfait_reaching_2016,bienfait_magnetic_2017}, and the search for axions \cite{brubaker_first_2017}.
As the first component of a microwave amplification chain, the main desired specifications for a Josephson amplifier are: 
(i) low added noise: the noise added by the amplifier should be no larger than the minimum imposed by quantum mechanics, 
(ii) high gain: the amplifier power gain should be large enough to overwhelm the noise temperature of the following amplification chain (in practice, at least 20 dB), 
(iii) large bandwidth: the amplifier gain should be constant over a bandwidth that is large enough for the desired application, 
(iv) large dynamic range: the output signal power should be linearly proportional to the input signal power over a wide enough power range, 
(v) unidirectionality: the amplifier should, ideally, amplify only signals incident from the system being probed and isolate the signal source from spurious noise that propagates back from subsequent devices in the amplification chain,
(vi) ease of operation: the energy necessary for amplification should be delivered to the amplifier in a simple and robust manner without requiring precise tuning, 
(vii) robustness of construction: the amplifier circuit should not require too delicate tolerances.

Among these characteristics, dynamic range is a particularly important requirement for scaling up superconducting qubit setups to larger size systems \cite{bravyi_quantum_1998}. The dynamic range characterizes the input power range over which the amplifier behaves as a linear device for a single-tone or multitone input. For quantum-limited amplifiers, the lower limit on the dynamic range is set fundamentally by quantum mechanics, so improving dynamic range corresponds to increasing the upper limit.
The upper limit is controlled by two distinct but closely related nonidealities in the large-signal amplifier response. The first nonideality is the phenomenon of amplifier saturation, also called gain compression. This limits the maximum output power that can be produced by the device for an arbitrary input signal.
The second nonideality, previously unexplored for quantum-limited amplifiers, is the phenomenon of intermodulation distortion for multitone inputs, where the amplifier produces spurious tones on its output in addition to the desired amplified copies of the input tones.
Together, these two nonidealities limit the signal powers that can be processed by the amplifier and thus are a problem for faster or higher-power qubit readout as well as for the readout of multiple qubits \cite{jeffrey_fast_2014}.


Is it possible to improve the amplifier dynamic range without sacrificing other desirable characteristics? 
Here, we answer affirmatively by demonstrating systematic improvement of the dynamic range of a 3-wave-mixing degenerate parametric amplifier, named the SNAIL Parametric Amplifier (SPA). The SPA is based on an array of Superconducting Nonlinear Asymmetric Inductive eLements (SNAILs) \cite{frattini_3-wave_2017,zorin_josephson_2016}, which provides the flexibility needed to optimize the 3-wave-mixing amplification process, 
while simultaneously minimizing the 4-wave-mixing Kerr nonlinearity suspected to cause amplifier saturation \cite{eichler_controlling_2014,liu_josephson_2017}. With this flexibility, we have engineered an SPA that achieves a 1 dB compression power ($P_{-1\rm{dB}} \in [-102, -112]$ dBm for 20 dB gain) on par with the best quantum-limited resonant parametric amplifiers \cite{mutus_strong_2014,eichler_quantum-limited_2014,roy_broadband_2015}, but over the entire tunable bandwidth of $1 \,\text{GHz}$ without sacrificing any other desirable characteristics, including quantum-limited noise performance.

Our demonstration of dynamic range improvement is crucially accompanied by first-principles theory that elucidates the link between the physical realization of the amplifier and the nonidealities of its response to large input signals. This link is accomplished in two steps. First, we show how to map the physical layout of the SPA to the phenomenological parameters that enter the input-output description of the device. These parameters consist of the 3- and 4-wave-mixing nonlinear components of the SPA Hamiltonian, as well as the damping induced through coupling to a transmission line via an input-output port. Second, we describe and validate experimentally how these phenomenological parameters directly determine the nonidealities in the amplifier's response to large input signals. Such first-principles theoretical description opens the door to further improvements in the amplifier dynamic range as well as the optimization of any other form of parametrically induced mixing for quantum information processing.

The article is organized as follows. In Sec.~\ref{sec:physicalKnobs}, we introduce the physical realization of the SPA. Sec.~\ref{sec:SPAmodel} briefly describes the relevant parameters of the SPA model and their influence on the small-signal gain of the amplifier. Sec.~\ref{sec:measureH} validates our mapping between the physical layout and the SPA parameters, with the theoretical details given in Appendix \ref{sec:app_Hspa}. In Sec.~\ref{sec:GainCompression}, we explore the mechanisms responsible for amplifier saturation, and characterize the intermodulation distortion of the SPA in Sec.~\ref{sec:Intermods}, with theoretical details given in Appendix \ref{sec:app_DR}.

\begin{figure}
 \includegraphics[angle = 0, width = \figwidth]{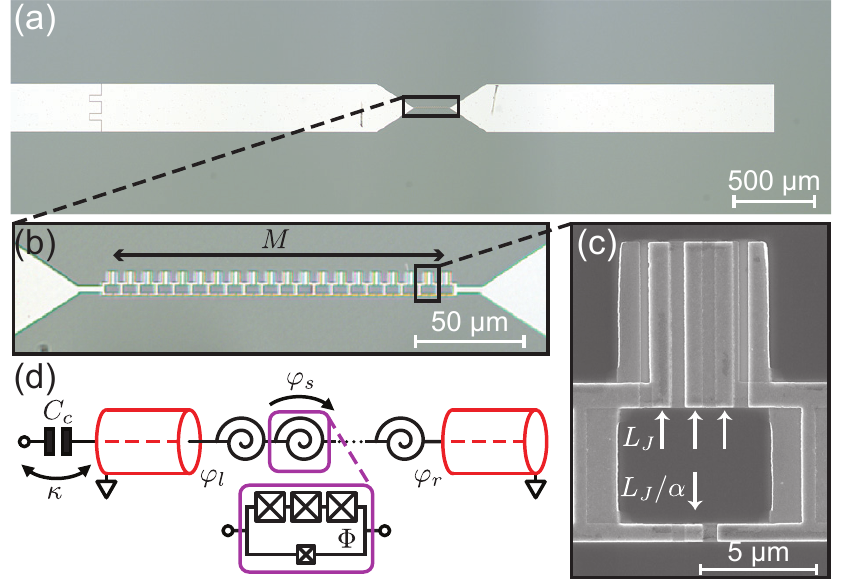}
 \caption{\label{fig1}  (a) Optical microscope image and (d) corresponding circuit model of a SNAIL Parametric Amplifier (SPA). An array of $M$ SNAILs is inserted at the center of a $\lambda/2$ section of microstrip transmission line, colored red in (d). (b) Image of an array of $M=20$ SNAILs. (c) Electron micrograph of a single SNAIL with 3 large Josephson junctions (inductance $L_J$) in a loop with one smaller junction (inductance $L_J/\alpha$). Arrows indicate the junctions and the inset of (d) gives the SNAIL circuit schematic. In (d), $\varphi_s$ denotes the phase drop across each SNAIL. The node phase $\varphi_l$ ($\varphi_r$) denotes the location where the left (right) side of the array of SNAILs connects to the linear embedding structure. The dissipation rate $\kappa$ is set by capacitive coupling with capacitance $C_c$ to the transmission line.}
\end{figure}
%
%
\begin{table}
\begin{ruledtabular}
\begin{tabular}{c | c | c | c | c | c}
Device & $L_J$ (pH)& $M$ & $\alpha$ & $C_c$ (pF)  & $\omega_0/2\pi$ (GHz) \\
\hline
A	& 60 & 1 & 0.29 & 0.048 & 8.4	 \\
B 	& 67 & 10 & 0.29 & 0.039 & 11.4	\\
C	& 47 & 20 & 0.09 & 0.068 & 17.9	\\
D	& 44 & 20 & 0.09 & 0.075 & 23.5	\\
E	& 34 & 20 & 0.09 & 0.088 & 23.4
\end{tabular}
\end{ruledtabular}
\caption{\label{tab:devs} Constitutive parameters of 5 devices measured in the experiment: Josephson inductance of largest junction ($L_J$), number of SNAILs ($M$), junction \blue{inductance} ratio ($\alpha$), coupling capacitance to the $50 \,\Omega$ transmission line ($C_c$), and frequency of the $\lambda/2$ microstrip embedding structure when the array of SNAILs is replaced by a short ($\omega_0$).}
\end{table}
%
%
\section{SPA Physical Realization} \label{sec:physicalKnobs}
Akin to the Josephson Parametric Amplifier (JPA) \cite{castellanos-beltran_amplification_2008,eichler_observation_2011}, the SPA is realized by placing $M$ SNAILs at the center of a $\lambda/2$ section of microstrip transmission line (Fig.~\ref{fig1}a). Fig.~\ref{fig1}b depicts an array of $M=20$ SNAILs, where each SNAIL consists of an array of 3 large Josephson junctions (Josephson inductance $L_J$) in a loop with one smaller junction (inductance $L_J/\alpha$). In practice, we chose the smallest $L_J$ that was still larger than the parasitic geometric inductance of the $24 \,\mu\text{m}$ perimeter SNAIL loop. As shown in the electron micrograph (Fig.~\ref{fig1}c), the Josephson junctions are fabricated using a Dolan bridge process for aluminum (Al) on silicon (Si).

The microstrip transmission line sections are formed by a 2 $\mu\text{m}$ thick silver (Ag) layer deposited on the back of the $300 \,\mu\text{m}$ thick Si wafer to act as a ground plane, and by center traces of Al whose length $l_{\rm{MS}}$ and width $w_{\rm{MS}}$ adjust the frequency $\omega_0$ and the characteristic impedance $Z_c$. For all devices in this work, we held the microstrip width constant at $w_{\rm{MS}} = 300 \,\mu\text{m}$ to set $Z_c= 45 \,\Omega$, and adjusted $l_{\rm{MS}}$ (in conjunction with $M$, $\alpha$ and $L_J$) to set the operating frequencies of the devices (see Section \ref{secsub:linear}). The coupling to the $50 \,\Omega$ transmission line $\kappa$ is set by a gap capacitor (capacitance $C_c$) at one end of the SPA resonator (Fig.~\ref{fig1}a). Later devices (E in Table \ref{tab:devs}) also have a second weakly capacitively coupled port on the opposite end of the resonator for the delivery of the pump (not shown in Fig.~\ref{fig1}a). By design, $\kappa$ is much larger than the internal dissipation rate and the coupling to the pump port.

The experimental characterization was performed in a helium dilution refrigerator (temperature $\approx 20 \,\text{mK})$ with a standard microwave reflection measurement setup. 
While cold, a magnet coil mounted beneath the sample applies a magnetic flux $\Phi$ to each SNAIL, which we assume to be uniform across the array. All measurements were performed with a PNA-X network analyzer \footnote{PNA-X network analyzer model Keysight N5242A.}, which contains two microwave sources and the capability to quickly perform intermodulation distortion measurements (see Section \ref{sec:Intermods}). The strong pump tone needed for amplification was either combined with the signal tone at room temperature or applied on a separate pump line.

\section{SPA Model} \label{sec:SPAmodel}
The device is modeled with the circuit schematic of Fig~\ref{fig1}d. Following Ref.~\onlinecite{frattini_3-wave_2017}, we treat the SNAIL as a nonlinear inductor that provides an asymmetric potential energy $U_{\rm{SNAIL}}(\varphi_s)$ and corresponding current-phase relation $I_s(\varphi_s) =  \frac{2\pi}{\Phi_0} \frac{dU_{\rm{SNAIL}}}{d\varphi_s}$ (where $\Phi_0 = h/2e$ is the superconducting magnetic flux quantum and $\varphi_s$ is the phase drop across the small junction of the SNAIL). These functions are engineered via the junction inductance ratio $\alpha$ and the externally applied magnetic flux $\Phi$. To include the linear embedding circuit, we enforce the constraint of current conservation at the left and right boundary nodes of the SNAIL array (phases denoted $\varphi_l$ and $\varphi_r$ in Fig.~\ref{fig1}d), which are connected to the ends of the respective transmission lines. As shown in Appendix \ref{sec:app_Hspa}, properly handling this nonlinear constraint equation is crucial for the prediction of higher order Hamiltonian terms, such as Kerr.

We next quantize the system and express the Hamiltonian of the lowest frequency mode of the SPA up to fourth order as 
\begin{equation}
\label{Hspa}
\bm H_{\rm{SPA}} / \hbar = \omega_a \bm{a}^{\dagger} \bm{a} + g_3 \left(\bm{a} + \bm{a}^{\dagger}\right)^3 + g_4 \left(\bm{a} + \bm{a}^{\dagger}\right)^4,
\end{equation}
where $\bm{a}^{\dagger}$ ($\bm{a}$) is the harmonic oscillator creation (annihilation) operator, $\omega_a$ is the resonant frequency, and the third-order and fourth-order nonlinearities are denoted $g_3$ and $g_4$ respectively. These three Hamiltonian terms are all tuned in situ via the applied magnetic flux $\Phi$ through each SNAIL loop. Along with the coupling rate to the transmission line $\kappa$, the parameters of $\bm H_{\rm{SPA}}$ determine the behavior of the SPA as a degenerate parametric amplifier as we show next.

To operate the SPA as a 3-wave-mixing amplifier, we apply a strong microwave pump tone at $\omega_p \approx 2\omega_a$, with mean intracavity amplitude $\alpha_p$. As shown in Appendix \ref{sec:app_DR}, input-output theory \cite{clerk_introduction_2010} gives  the phase-preserving power gain $G$ for a signal at frequency $\omega_s$ scattering in reflection off of an SPA as
\begin{equation}
\label{gain_simple}
G=1+\frac{4\kappa^{2}|g|^{2}}{(\Delta_{p}^{2}-\omega^{2}+\frac{\kappa^{2}}{4}-4|g|^{2})^{2}+(\kappa\omega)^{2}},
\end{equation}
where $\omega = \omega_s - \omega_p/2$ is the detuning of the input signal from $\omega_p/2$, $g = 2g_3 \alpha_p$, and $\Delta_{p} = \Delta + \frac{32}{3} g_4 |\alpha_{p}|^{2}$ with $\Delta = \omega_a - \omega_p / 2$.

For this work, we always set the pump frequency so that $\Delta = 0$.
Note that the maximum of gain $G$ always occurs at $\omega_s = \omega_p/2$, similar to the flux-pumped JPA \cite{yamamoto_flux-driven_2008}, making this amplifier particularly easy to tune up and operate. 
This property is in contrast with the tune-up procedure for Josephson Parametric Converters (JPCs) and even 4-wave-mixing JPAs, as outlined in Ref.~\onlinecite{liu_josephson_2017} and Ref.~\onlinecite{hatridge_dispersive_2011} respectively.

As shown by Eq.~\ref{gain_simple}, designing an amplifier operating at $\omega_s$ reduces to engineering $\omega_a$, $g_3$, $g_4$, and $\kappa$. This task is accomplished by the appropriate choice of the physical knobs described in Section \ref{sec:physicalKnobs}. To illustrate control over these Hamiltonian parameters and provide intuition on this mapping, we compare the set of devices listed in Table \ref{tab:devs}.
\begin{figure}
\includegraphics[angle = 0, width = \figwidth]{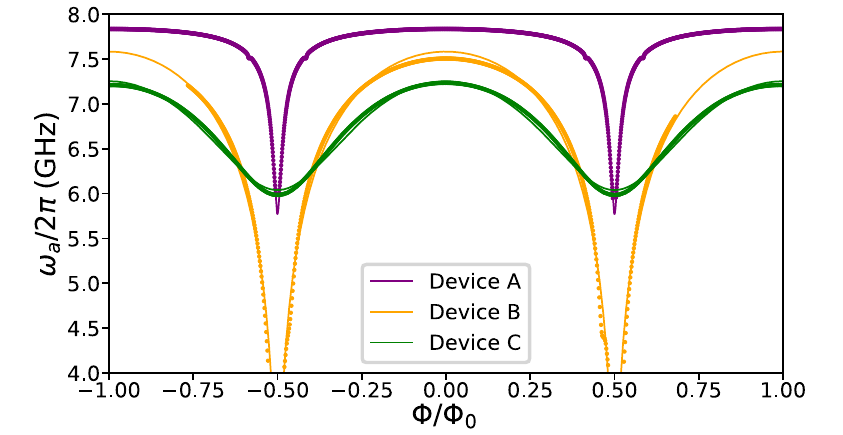}
\caption{\label{fig2}   Resonant frequency $\omega_a$ as a function of applied magnetic flux $\Phi$ for three devices. Thin solid lines are fits to a model based on the schematic in Fig.~\ref{fig1}d.}
\end{figure}
\section{SPA Hamiltonian Characteristics} \label{sec:measureH}
\subsection{Resonant Frequency Tunability} \label{secsub:linear}
We first compare the linear-response characteristics of these devices, specifically the resonant frequency $\omega_a$ as a function of applied magnetic flux $\Phi$ (Fig.~\ref{fig2}). The tunability range of $\omega_a$ depends on two factors: (1) the flux-tunable SNAIL inductance $L_{\text{SNAIL}}(\Phi)$, and (2) the participation of the SNAIL array in its embedding structure. Multiple physical knobs affect both of these factors; here for simplicity we focus on the influence of $\alpha$ and $M$.

The first factor, the flux dependence of $L_{\text{SNAIL}}(\Phi)$, is strongly analogous to that of a dc superconducting quantum interference device (SQUID) or an rf SQUID \cite{clarke_squid_2004}: the inductance is tunable between a minimum at $\Phi/\Phi_0 = 0$ and maximum at $\Phi/\Phi_0 = 0.5$. The range of this tunability is given by the asymmetry between the inductances on either arm of the superconducting loop which, in the SNAIL, is controlled by the junction inductance ratio $\alpha$. $\alpha = 1/3$ (where 3 is the number of large junctions in a SNAIL) corresponds to perfect inductive symmetry, resulting in $L_{\text{SNAIL}}(0.5 \, \Phi_0) \to \infty$. $\alpha > 1/3$ causes the SNAIL potential to have multiple inequivalent minima and results in hysteretic behavior, which we wish to avoid in an amplifier. $\alpha < 1/3$ gives some asymmetry, where smaller $\alpha$ corresponds to a smaller inductive tunability range. However, as we will show in Section \ref{secsub:nonlinear}, smaller $\alpha$ is advantageous for achieving the optimal flux profile of $g_3$ and $g_4$.

The second factor influencing the tunable range of $\omega_a$ is the fraction of the mode inductance coming from the SNAILs. For a given SNAIL design with an $L_{\text{SNAIL}}(\Phi)$, this is controlled by the number of SNAILs $M$ in series as well as the length $l_\text{MS}$ and width $w_\text{MS}$ of the surrounding microstrip embedding structure. In practice, $M$ provides more control over $\omega_a$ due to the practical difficulty in realizing microstrip embedding structures with impedances significantly different from 50 $\Omega$. Thus, $M$ and $\alpha$ are chosen first and then we adjusted $l_{\text{MS}}$ (while keeping $w_\text{MS}$ fixed) to hit our desired operating frequency range.

Focusing on the $\Phi$ dependence of $\omega_a$ in devices B and C (Fig.~\ref{fig2}), we see the ability of $\alpha$ and $M$ to engineer the frequency tunability. For $\alpha = 0.29$ as in Device B, the total inductances on either arm of the SNAIL are nearly equal so the SNAIL inductance changes drastically from $\Phi/\Phi_0 = 0$ to $\Phi/\Phi_0 = 0.5$. Conversely, the inductance of each $\alpha=0.09$ SNAIL in device C changes only a little and the aggregation of these small changes for all 20 SNAILs gives the device its approximately 1 GHz of tunability.
\begin{figure}
\includegraphics[angle = 0, width = \figwidth]{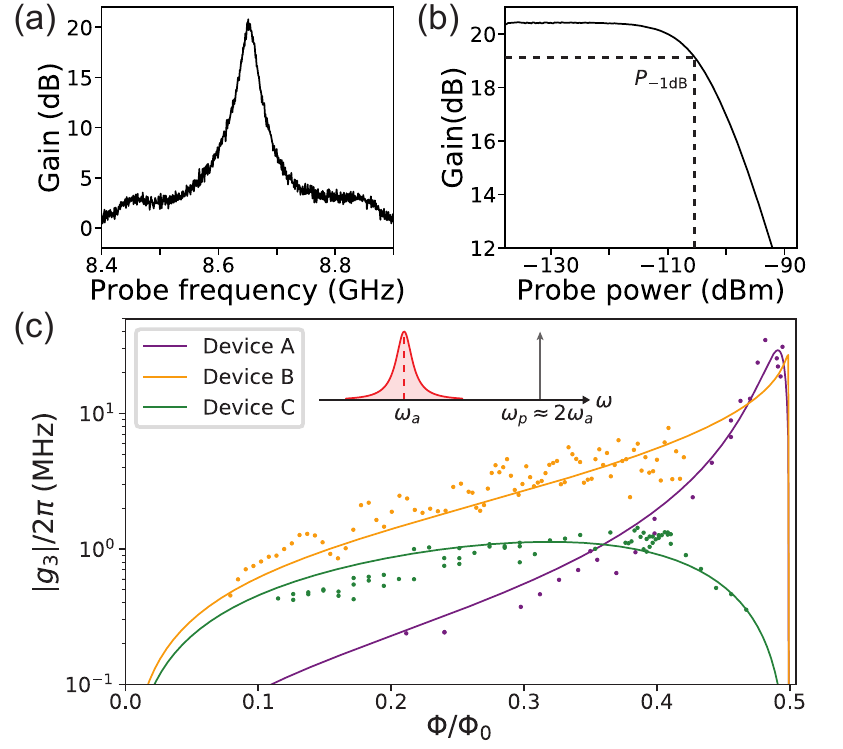}
\caption{\label{fig3}   (a) Reflection gain versus input signal frequency when the SPA is biased with a strong tone at $\omega_p \approx 2 \omega_a$, frequency landscape inset of (c). (b) Gain as a function of input signal power shows amplifier saturation. Input power at which the gain reduces by 1 dB is denoted $P_{-1\textrm{dB}}$. (c) Third-order nonlinearity $g_3$ versus applied magnetic flux $\Phi$. Solid curves are first-principles theory for $g_3$.}
\end{figure}
\subsection{Nonlinear Characteristics} \label{secsub:nonlinear}
Having described the linear response of the SPA resonator, we next demonstrate its operation as a 3-wave-mixing degenerate parametric amplifier. We applied a strong microwave pump tone at $\omega_p = 2 \omega_a$ and adjusted the pump power to achieve 20 dB of small-signal reflection gain (example shown Fig.~\ref{fig3}a).
One standard phenomenon that limits amplifier quality is the saturation of the gain with increasing input signal power. Shown in Fig.~\ref{fig3}b, we measured the input-referred 1 dB compression point $P_{-1\rm{dB}}$ as the input signal power where the gain drops by 1 dB. 
To understand this phenomenon and improve the $P_{-1\rm{dB}}$, we perform a systematic study across multiple devices.
Toward this goal, we begin by first measuring the nonlinearity of the SPA Hamiltonian (Eq.~\ref{Hspa}) as a function of applied magnetic flux $\Phi$ for all devices in Table \ref{tab:devs}.

The dependence of third-order nonlinearity $g_3$ on $\Phi$ is shown in Fig.~\ref{fig3}c for three representative devices. $g_3$ is extracted from Eq.~\ref{gain_simple} by tuning up a 20-dB gain point and using the measured values for $\omega_a$, $\kappa$ and a calibration on the applied pump power. Also shown is our first-principles theory calculation, which uses only the linear characteristics fit from Fig.~\ref{fig2} and room-temperature measurements of the resistance of the SNAIL array. A global scale factor of $\approx 2$ has been applied to the extracted $g_3$, which could arise from pump-power miscalibration or the enhanced coupling of the pump to the SNAILs through higher-frequency modes not considered in our simple model. Comparing devices A and C, we note the relatively constant $g_3$ for device C ($\alpha=0.09$) except near $\Phi/\Phi_0 = 0$ and $\Phi/\Phi_0 = 0.5$ where symmetry forbids 3-wave-mixing terms. In contrast, device A ($\alpha=0.29$) shows a two order of magnitude variation in $g_3$ over the same flux range. This comparison highlights the drastic difference in the flux profile of $g_3$, here mainly arising from the difference in the junction inductance ratio $\alpha$.

\begin{figure}
\includegraphics[angle = 0, width = \figwidth]{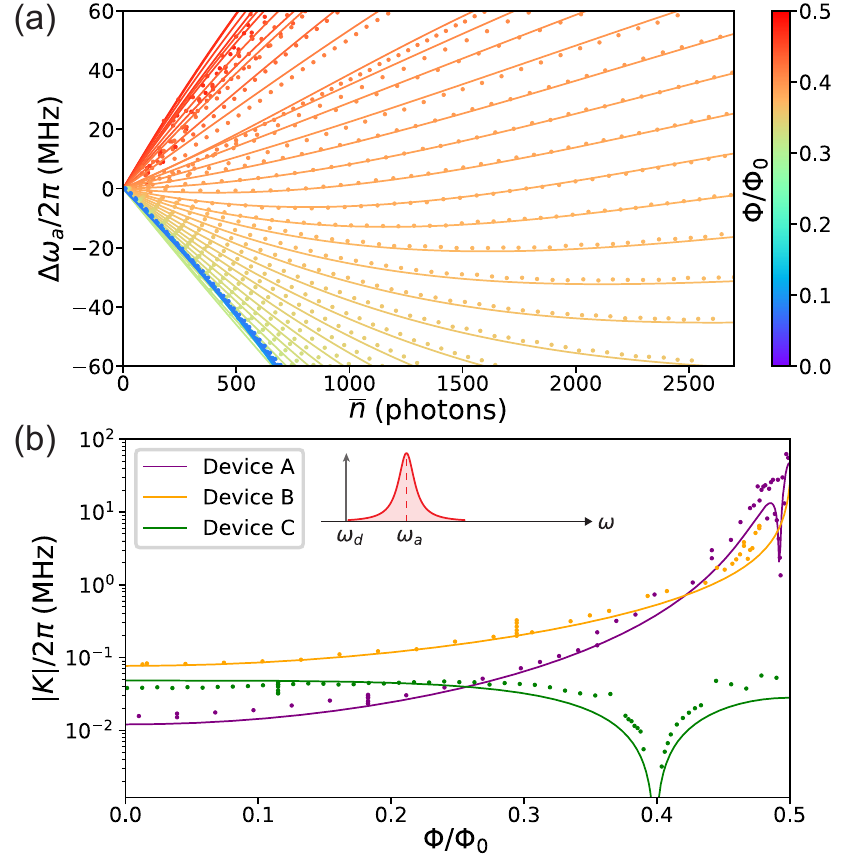}
\caption{\label{fig4_g4}   (a) Frequency shift $\Delta\omega_a$ versus the number of steady-state photons populating the resonator $\bar{n}$ induced by a drive at $\omega_d$ (inset of (b)).
The measured shift is plotted for a few different applied magnetic fluxes $\Phi$ (denoted with color). Solid lines are fits to $\Delta\omega_a = 2 K\bar{n} + K' \bar{n}^2$. (b) Magnitude of Stark shift per photon $|K|$ as a function of applied magnetic flux $\Phi$. Solid lines are first-principles theory for $|K|$.}
\end{figure}

The fourth-order nonlinearity $g_4$ is extracted from a Stark shift measurement. In this experiment, we applied a strong $\approx 500$ MHz detuned drive that populates the resonator with $\bar{n}$ average steady-state photons and shifts its resonant frequency. Here, $\bar{n}$ is calibrated using fits of $\omega_a$, $\kappa$ and room-temperature line attenuation. In Fig.~\ref{fig4_g4}a, we plot the measured frequency shift $\Delta\omega_a$ of a typical SPA resonator as a function of $\bar{n}$ and applied magnetic flux $\Phi$ (color). The frequency shift changes from negative to positive over half of a flux quantum. The solid lines are fits to $\Delta\omega_a = 2 K\bar{n} + K' \bar{n}^2$.
From this fit, we extract the Stark shift per photon $K$, which is related to the Hamiltonian parameters $g_3$ and $g_4$ up to second order in perturbation theory by $K = 12 (g_4 - 5g_3^2/\omega_a)$ (see Appendix \ref{sec:app_Hspa}).

The dependence of $K$ and thus $g_4$ on $\Phi$ is shown in Fig.~\ref{fig4_g4}b for three representative devices together with our first-principles theory calculation. The contrast between device A and device C again highlights the effect of $\alpha$ on the flux profile. Specifically, device A shows a three order of magnitude change in $g_4$, while device C's $g_4$ is relatively constant over most of the flux range. Additionally, both devices nominally support a region of suppressed Kerr. However, device A attains this region over a very narrow flux range, making the suppression practically useless, while device C shows a robust suppression regime by more than an order of magnitude from its $\Phi/\Phi_0 = 0$ value. This suppression could be useful in applications where the circuit designer wants some nonlinearity for mixing purposes, but would prefer to suppress spurious Kerr interactions.

While our previous comparison of devices A-C focused on the flux profile of $g_3$ and $g_4$, their overall magnitudes must also be engineered for optimizing amplifier nonidealities, such as saturation power. Besides $\alpha$ and $L_J$, these magnitudes are also influenced by the number of SNAILs $M$ (see Appendix \ref{sec:app_Hspa}). At small values, changing $M$ strongly affects $|g_3|$ and $|g_4|$, but we note its influence substantially weakens for $M \gtrsim 20$.
Subsequent devices (D and E) have similar magnitudes of the nonlinearities and flux profiles to device C, but instead vary the coupling to the transmission line $\kappa$. A summary of these phenomenological parameters for all devices is given in Table \ref{tab:Hparams}. As we show next, these factors affect the gain compression.

%
%
\begin{table}
\begin{ruledtabular}
\begin{tabular}{c | c | c | c | c}
Device & $\omega_a/2\pi$ (GHz) & $\kappa/2\pi$  & $|g_3|/2\pi$ & $|g_4|/2\pi$ \\
\hline
A	& <6 - 7.84 & 35 - 55 & 0.3 - 30	&	0.001 - 4.9	 	\\
B 	& <4 - 7.51 & 30 - 35 & 0.5 - 60	&	0.006 - 0.5		\\
C	& 5.99 - 7.24 & 90 - 120 & 0.4 - 1.5	&	0.004	\\
D	& 7.09 - 8.37 & 180 - 250 & 0.5 - 1.8	&	0.003	\\
E	& 7.76 - 9.24 & 270 - 440 & 0.7 - 2.0	&	0.004
\end{tabular}
\end{ruledtabular}
\caption{\label{tab:Hparams} \textit{In situ} tunable range of phenomenological parameters of five devices measured in the experiment: resonant frequency ($\omega_a$), coupling to the $50 \,\Omega$ transmission line ($\kappa$), third-order nonlinearity ($g_3$), and fourth-order nonlinearity ($g_4$) where we quote the average for devices C, D, and E disregarding the $0.1 \,\Phi_0$ region around the Kerr-free point. All parameters given in MHz, except for $\omega_a/2\pi$ in GHz.}
\end{table}

\section{Gain Compression} \label{sec:GainCompression}
Having established the connection between the physical parameters of our device and the properties of $\bm H_{\rm{SPA}}$ (Eq.~\ref{Hspa}), we now optimize the nonlinearities ($g_3$ and $g_4$) and the coupling to the transmission line ($\kappa$) to achieve higher dynamic range. But first, let us review the causes of amplifier saturation.

The previous formula for gain (Eq.~\ref{gain_simple}) shows no dependence on input signal power, and therefore does not capture the phenomenon of amplifier saturation. To include this dependence, we need to account for the population of the resonator by signal photons at frequency $\omega_s$. We therefore introduce the mean intracavity amplitude $\alpha_s$. Furthermore, 3-wave mixing creates an image tone (often called the idler)  at frequency $\omega_i = \omega_p - \omega_s$, with intracavity amplitude $\alpha_i$, which is comparable to $\alpha_s$. As shown by a semiclassical harmonic-balance analysis (see Appendix \ref{sec:app_DR}) that includes both of these amplitudes in the input-output theory on equal footing with $\alpha_p$, the gain $G$ can be recast into a formula similar to Eq.~\ref{gain_simple}. We find
\begin{equation}
\label{gain}
G=1+\frac{4\kappa^{2}|g_{\rm{eff}}|^{2}}{(\Delta_{\rm{eff}}^{2}-\omega^{2}+\frac{\kappa^{2}}{4}-4|g_{\rm{eff}}|^{2})^{2}+(\kappa\omega)^{2}},
\end{equation}
where $\omega = \omega_s - \omega_p/2$, $g_{\rm{eff}} = 2g_3 \alpha_p$, and $\Delta_{\rm{eff}}=\Delta + 12g_{4}\left[\frac{8}{9}|\alpha_{p}|^{2} +|\alpha_{s}|^{2} +|\alpha_{i}|^{2}\right]$ with $\Delta = \omega_a - \omega_p / 2$. Considering the on resonance response ($\omega \to 0$), we see that we can tune the pump strength and thus $g_{\rm{eff}}$ such that the denominator of Eq.~\ref{gain} goes to 0 and the gain $G$ diverges. Resonant parametric amplifiers operate very close to this parametric instability point, with $g_{\rm{eff}}$ chosen such that $G = 20$ dB. As a result, slight changes in this denominator are enough to significantly affect the gain $G$.

Two causes of gain compression can be associated with changes in the denominator of Eq.~\ref{gain}. The first, Kerr-induced Stark shifts, comes from shifts in $\Delta_{\rm{eff}}$ with increasing signal power. More signal power increases $\alpha_s$ and shifts the resonant frequency due to $g_4$. Under the approximation that $\alpha_p$ is independent of $\alpha_s$ (often termed the stiff-pump approximation \cite{kamal_signal--pump_2009,roy_quantum-limited_2018}), we can calculate the compression power due to Stark shifts as
\begin{equation}
P_{-1\text{dB}}^{\text{Stark}}\sim \frac{\kappa}{|g_{4}|} \frac{1}{G_0^{5/4}} \hbar \omega_a \kappa, \label{eq:p1dB Stark}
\end{equation}
where $G_0$ is the small-signal gain (see derivation in Appendix \ref{sec:app_DR}).

The second cause of gain compression visible from Eq.~\ref{gain} is that of pump depletion, which arises from the breakdown of the stiff-pump approximation \cite{kamal_signal--pump_2009,roy_quantum-limited_2018}. Pump depletion results from the intrinsic nonlinear coupling between the intracavity pump amplitude $\alpha_p$ and the signal amplitude $\alpha_s$. Thus, increasing $\alpha_s$ changes $\alpha_p$ and consequently the denominator of Eq.~\ref{gain}. Assuming $g_4 = 0$, we can estimate the compression power due to pump depletion as
\begin{equation}
P_{-1\text{dB}}^{\rm{pump \,dep}}\sim \frac{\kappa}{g_{3}^{2} / \omega_a}\frac{1}{G_0^{3/2}} \hbar \omega_a \kappa, \label{eq:p1dB pumpDep}
\end{equation}
where $G_0$ is the small-signal gain. We note that this compression mechanism arises directly from the third-order nonlinearity that we need for amplification, and is thus unavoidable.

Given these limits on dynamic range, we examine Eq.~\ref{eq:p1dB Stark} and Eq.~\ref{eq:p1dB pumpDep} to formulate a recipe for higher compression powers: decrease nonlinearities $g_3$ and $g_4$, and increase the dissipation $\kappa$. Intuitively, this recipe pushes the optimization closer to a system that obeys the assumptions underlying Eq.~\ref{gain_simple}, namely a more linear oscillator.

Following this recipe requires more applied pump power to reach a desired gain. However, we must be mindful that the current through the SNAIL does not approach the critical current of its Josephson junctions. In practice, applying pump currents that approach the critical current does not directly cause gain compression, but instead determines whether the amplifier achieves the desired small-signal gain in the first place. This limitation translates  to $pQ \gtrsim 1$, where $p$ is the inductive participation ratio of all nonlinear elements and $Q$ is the total quality factor of the SPA mode \cite{eichler_quantum-limited_2014}. This is also rigorously equivalent to ensuring the validity of the Taylor expansion of the SNAIL potential in deriving $\bm H_{\rm{SPA}}$ (Eq.~\ref{Hspa}). All amplifiers we consider here satisfy $pQ > 15$ to ensure that the amplifier produced 20 dB of small-signal gain.

\begin{figure}
\includegraphics[angle = 0, width = \figwidth]{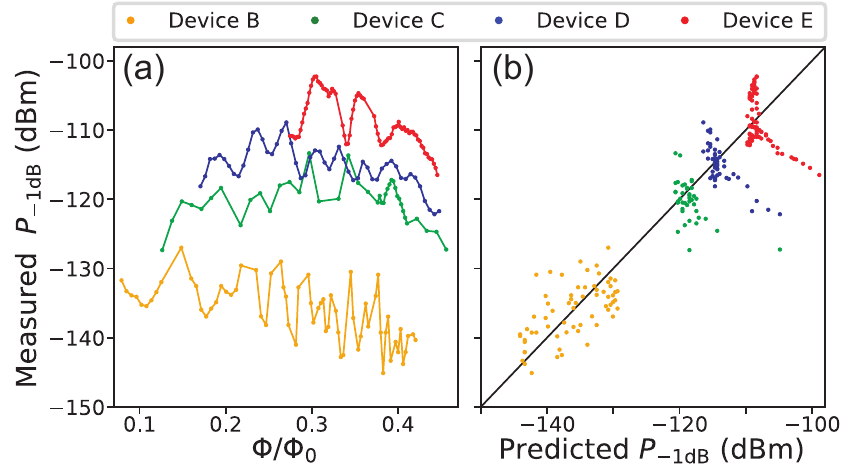}
\caption{\label{fig5_P1dB}   (a) Measured 1-dB compression power ($P_{-1\rm{dB}}$) as a function of applied magnetic flux $\Phi$ for four devices biased at 20-dB gain. (b) Comparison between measured value and first-principles theory, which semiclassically treats pump depletion and Stark shifts to second order in harmonic balance.}
\end{figure}
%
We followed the recipe of reducing nonlinearities and increasing dissipation in designing our devices (see Table \ref{tab:Hparams}), and compare their 1-dB compression powers in Fig.~\ref{fig5_P1dB}a as a function of applied flux $\Phi$. For each point, we measured the resonant frequency $\omega_a$, applied a pump at $\omega_p = 2\omega_a$, and adjusted its power to get $G= 20$ dB. We then measured the $P_{-1\rm{dB}}$ compression point (as shown in Fig.~\ref{fig3}b). Fig.~\ref{fig5_P1dB}b shows the correlation of our first-principles theory predictions of saturation power with the measured $P_{-1\rm{dB}}$, where the black line indicates agreement between theory and experiment. This theory numerically solves the semiclassical Langevin equations of motion to second order in harmonic balance to obtain $\alpha_s$, $\alpha_i$, and $\alpha_p$ for given input pump and signal powers (see Appendix \ref{sec:app_DR}). The gain is then calculated using Eq.~\ref{gain}. We find that, for our devices, the Stark shift mechanism of gain compression closely approximates the full numerical solutions.

To confirm the dependence of $P_{-1\rm{dB}}$ on $g_4$, we first focus on device B, which has the largest $|g_4|/2\pi \in [6, 530]$ kHz for different flux bias points. This change in $g_4$ results in a systematic 15-dB change in $P_{-1\rm{dB}}$ and the theory predicts the trend. We note that the scatter in the data of Fig.~\ref{fig5_P1dB} results from ripples in the impedance of the transmission line seen by the SPA, which affects its linewidth $\kappa$. The compression power is highly sensitive to this parameter, which can be seen in Eq.~\ref{eq:p1dB Stark} and Eq.~\ref{eq:p1dB pumpDep}.

For device C, we engineered $\kappa/2\pi \in [90, 120]$ MHz and  $|g_4|/2\pi \approx 4$ kHz except near its Kerr-free region (see Fig.~\ref{fig4_g4}b). These changes in $\kappa$ and $g_4$ directly result in device C's increased performance compared to device B. Devices D and E are similar to device C but with increasing $\kappa/2\pi$ to $[180, 250]$ MHz and $[270, 400]$ MHz respectively, and again show improved performance. Specifically, the best device, device E, achieves $P_{-1\rm{dB}} \in [-102, -112]$ dBm, which is on par with the best known quantum-limited resonant parametric amplifiers \cite{mutus_strong_2014,eichler_quantum-limited_2014,roy_broadband_2015}. We stress that this performance, achieved with a dynamic bandwidth $\approx 30-40$ MHz, is consistent over the entire tunable bandwidth of 1 GHz.

Despite this increase in dynamic range, Fig.~\ref{fig5_P1dB}b shows that theory predicted that we should have achieved higher saturation powers at certain applied magnetic fluxes. The flux bias points where theory overpredicts $P_{-1\rm{dB}}$ are those where $g_4$ is suppressed (see Eq.~\ref{eq:p1dB Stark}). Specifically, device C in Fig.~\ref{fig4_g4}b shows a tenfold reduction in measured $g_4$ at around $\Phi/\Phi_0 = 0.4$. However, the measured $P_{-1\rm{dB}}$ did not increase near this flux bias point (Fig.~\ref{fig5_P1dB}a). Devices D and E show similar Kerr-free regions as measured by Stark shift, but also do not show increased $P_{-1\rm{dB}}$. This puzzle suggests that either our theory has misidentified the cause of amplifier compression, despite its success at all other flux points, or that Kerr is not, in fact, suppressed in  these regions when the pump is on and the amplifier is operational.

\begin{figure}
\includegraphics[angle = 0, width = \figwidth]{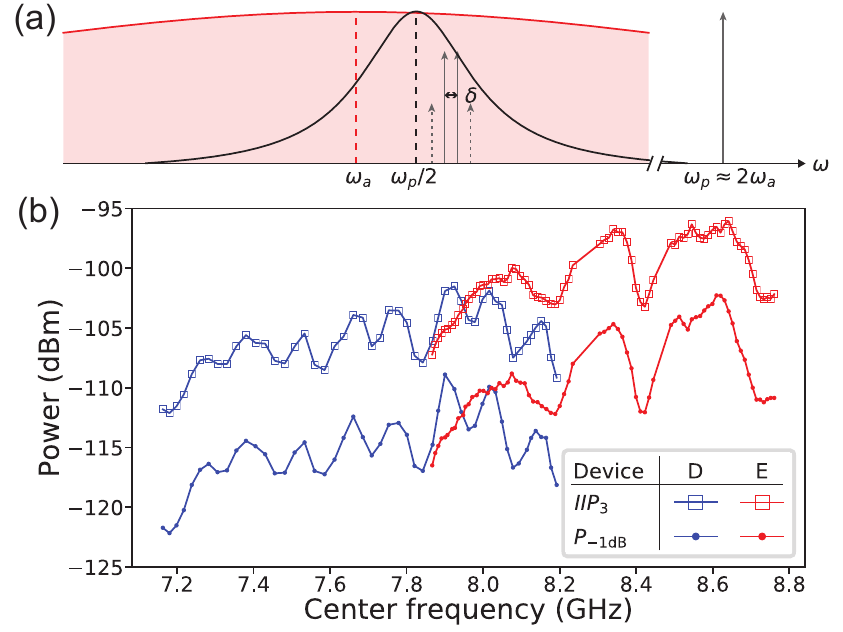}
\caption{\label{fig6_IMD}   (a) Caricatured (not-to-scale) frequency spectrum for the measurement of third-order intermodulation distortion products of an SPA. The red shaded region is the Lorenzian lineshape of the linear mode of width $\kappa$. Black is the reflection gain of the amplifier when pumped with a strong microwave tone at $\omega_p \approx 2\omega_a$. Solid arrows show two tones applied above $\omega_p / 2$ with a relative detuning $\delta$. Dashed arrows denote sidebands generated by the SPA detuned from the main tones by $\delta$. (b) $IIP_3$ and $P_{-1\rm{dB}}$ as a function of the center frequency ($\omega_p/2$) of the 20-dB gain curve for two devices. Neighboring experimental data points have been joined to emphasize correlations between the two experiments. }
\end{figure}
%
%
\section{Intermodulation Distortion} \label{sec:Intermods}
To investigate this discrepancy, we measured the Kerr nonlinearity in the presence of the strong pump tone using a third-order intermodulation distortion (IMD) measurement \cite{pozar_microwave_2012}. This standard nomenclature of third-order IMD originates from the fact that fourth-order Kerr terms in the Hamiltonian generate third-order terms in the equations of motion. As we will show, this measurement provides clues about the causes of amplifier saturation and also probes the response of the amplifier to multitone or broadband input signals. Understanding the response to such input signals is particularly crucial for employing quantum-limited amplifiers in any multiplexed readout scheme of superconducting qubits.

A third-order IMD experiment is performed according to the frequency landscape in Fig.~\ref{fig6_IMD}a. With the pump on and the amplifier biased to $G=20$ dB, we applied two main signal tones (solid gray arrows) centered at $\omega_p/2 + 2\pi \times 500 \,\text{kHz}$ with a relative detuning $\delta/2\pi = 100 \,\text{kHz}$ and measured the power in the resulting sidebands (dashed gray arrows). Intuitively, two signal photons from one input tone and one from the other combine in a 4-wave-mixing process to generate the resulting sideband. Thus, the measured relative power between the main tones and the sidebands indicates the amount of spurious 4-wave mixing occurring in the device.

Sweeping the applied power on the two main signal tones, we extracted the input-referred third-order intercept point ($IIP_3$) at which the measured sideband power would equal the main input signal power (details in Appendix \ref{sec:app_IMD} and Fig.~\ref{figS2_g4imd}a).
We conform to the usage in standard microwave amplifier data sheets to take $IIP_3$ as the metric for third-order IMD.
To characterize our amplifiers, in Fig.~\ref{fig6_IMD}b, we compare the $IIP_3$ and $P_{-1\rm{dB}}$ as a function of the center frequency of the Lorenzian gain curve for two different devices. Each point corresponds to a point tuned up in Fig.~\ref{fig5_P1dB}. Strikingly, the features in $P_{-1\rm{dB}}$, which are caused by ripples in the line impedance, are exactly reproduced in $IIP_3$. Such a comparison indicates that the cause of $IIP_3$, which is spurious 4-wave mixing, is most likely responsible for the saturation of the amplifier. This confirms Ref.~\onlinecite{liu_josephson_2017}'s assertion that Kerr is responsible for the saturation of state-of-the-art parametric amplifiers.

Quantitatively, lowest-order harmonic balance theory predicts (see Appendix \ref{sec:app_DR}) that the measured $IIP_3$ is related to the Kerr nonlinearity $g_4$ by the equation
\begin{equation} \label{eq:IIP3}
IIP_{3}=\frac{\kappa}{12|g_{4}|} \frac{1}{G_0^{3/2}} \hbar \omega_a \kappa,
\end{equation}
where $G_0$ is the small-signal gain. However, upon a closer examination of Fig.~\ref{fig6_IMD}b, we do not observe a distinct peak in $IIP_3$ for devices D and E, which both support regions where $g_4$ is suppressed. Thus, we see again as with $P_{-1\rm{dB}}$ that the nonlinear properties of the amplifier in the Kerr-free region (as measured by Stark shifts) did not show the expected improvement.

$IIP_3$ is a measure of nonlinear 4-wave-mixing scattering in the presence of the amplification pump, and thus gives us a clue as to the origin of this discrepancy near the Kerr-free region.
Such nonlinear scattering can arise from the multiple terms in the Hamiltonian (Eq.~\ref{Hspa}): for example, a $g_4$ process as well as, for instance, two cascaded $g_3$ processes. 
While for most of the flux range the $g_4$ process dominates, near the Kerr-free region $g_3$ is maximal and the cascaded processes become important.
Taking these processes into account is equivalent to going to higher order in harmonic balance, which is shown to improve the agreement between theory and experiment for $P_{-1\rm{dB}}$ (see Appendix \ref{sec:app_DR}).


Not only do IMD measurements help us understand the causes of amplifier saturation, they are also interesting in their own right since reduction of these spurious mixing processes is important for many applications. For instance, any scheme for multiplexed readout of superconducting qubits requires the independence of the readout channels. These spurious intermodulation products will directly limit the isolation between channels either by directly mixing them or by distorting pulses. Furthermore, such intermodulation products put an upper bound on the quantum efficiency of any practical amplifier since, without careful calibration, distortion of the incident quantum signal is unlikely to be accounted for in the experimentalist's demodulation scheme.

\section{Conclusion}
In conclusion, we have introduced the SPA, a 3-wave-mixing degenerate parametric amplifier, which is simple to design, fabricate, and operate.
Through a systematic study across multiple devices, we have confirmed that the fourth-order Kerr term in the amplifier Hamiltonian is the primary cause of gain compression and intermodulation distortion. With this insight, we have optimized the SPA to achieve 1-dB compression powers on par with the best reported values for resonant quantum-limited parametric amplifiers, but over the entire tunable bandwidth of 1 GHz of the device, and without sacrificing any other desirable characteristics.

Importantly, the most precious of these characteristics, quantum-limited noise performance, was confirmed through comparing noise-visibility-ratio (NVR) measurements. A proxy for noise temperature, NVR is the ratio between the noise power spectral density with the pump on and the pump off. All amplifiers in this work were measured to have comparable NVRs and thus comparable noise performances to other quantum-limited amplifiers measured in the same system~\cite{frattini_3-wave_2017,bergeal_phase-preserving_2010}. Moreover, an SPA was shown to improve the readout of a superconducting qubit in Ref.~\onlinecite{touzard_gated_2018}, where the quantum efficiency of a phase-sensitive measurement chain involving an SPA was measured to be $\eta = 0.6$ in a self-calibrated manner.

Our work on the improvement of amplifier performance can be carried out further. One puzzling observation was the absence of a peak in the saturation power at the Kerr-free point, despite the confirmation that Kerr-induced Stark shifts are the primary cause of gain compression. The IMD measurements suggested that Stark shifts caused by higher harmonics limit the saturation power at the Kerr-free point. A natural next step would be to understand how to reduce these spurious harmonics in the presence of a strong pump drive.
An alternative strategy would be to further reduce amplifier nonlinearity, which should increase the saturation power at the cost of requiring more pump power to achieve the same gain. As such, effectively engineering the pump-power delivery network to achieve the desired pump strength, without introducing excess noise or heating up the base plate of the dilution refrigerator, will become increasingly more crucial for higher dynamic range amplifiers. More broadly, the optimizations performed in this work for higher dynamic range do not conflict with recent approaches for enhanced dynamic bandwidth via impedance engineering \cite{roy_broadband_2015,naaman_high_2017}, nor with approaches for directional amplification \cite{abdo_directional_2013,sliwa_reconfigurable_2015,lecocq_nonreciprocal_2017,metelmann_quantum-limited_2014,metelmann_nonreciprocal_2015,ranzani_graph-based_2015}. The reduction of Kerr and the use of arrays of nonlinear elements should also increase the dynamic range of traveling wave amplifiers \cite{macklin_nearquantum-limited_2015,white_traveling_2015,vissers_low-noise_2016,zorin_josephson_2016,bell_traveling-wave_2015,zhang_josephson_2017}.

Furthermore, our results indicate that 3-wave mixing with an array of SNAILs is a particularly robust building block for information processing with superconducting quantum circuits. With this versatile tool, both the third-order and the Kerr nonlinear parameters can be controlled over many orders of magnitude. Moreover, the sign of Kerr can be changed and its magnitude suppressed \textit{in situ} by tuning an applied magnetic flux. Such control can be convenient in parameter regimes which are rather different from quantum-limited amplifiers, such as for instance, in superconducting qubits \cite{mundhada_generating_2017,puri_engineering_2017}. The method of arraying multiple SNAILs is more generally applicable for optimizing parametrically induced mixing, such as in quantum-limited switches \cite{naaman_-chip_2016,chapman_widely_2017}, frequency converters \cite{abdo_full_2013,allman_tunable_2014,flurin_superconducting_2015,pfaff_controlled_2017,axline_-demand_2018,kurpiers_deterministic_2018,campagne-ibarcq_deterministic_2018} and other quantum devices \cite{leghtas_confining_2015,vool_driving_2018}.

\section*{Acknowledgements}
We acknowledge fruitful discussions with Shantanu Mundhada and Luigi Frunzio. We are grateful to Vladislav Kurilovich and Pavel Kurilovich for performing Born-Oppenheimer elimination of the fast SNAIL mode (see Appendix \ref{sec:app_Hspa}) to verify our semiclassical model.
We also acknowledge the Yale Quantum Institute. Facilities use is supported by YINQE, and the Yale SEAS cleanroom. This research is supported by ARO under Grant No.~W911NF-14-1-0011 and No. W911NF-18-1-0212, and S.S. acknowledges support from ARO under Grant No. W911NF-14-1-0563.
\appendix
\section{SPA Hamiltonian} \label{sec:app_Hspa}
In this Appendix, we discuss the map between the physical layout of the SPA and the Hamiltonian parameters. Section \ref{lumped_model} describes a single-mode lumped-element model of the SPA and shows the importance of the nonlinear current conservation for predicting the Kerr nonlinearity. Section \ref{renormalization_K} provides further insight about the renormalization of Kerr by high-energy modes. In Section \ref{distributed_model} we treat the SPA resonator as a distributed circuit element and show the expressions referred to as first-principles theory in the main text.

\subsection{Lumped-element model of SPA \label{lumped_model}}
The SNAIL \cite{frattini_3-wave_2017} is a one-loop dipole element composed of three identical large Josephson junctions (inductance $L_J$) in one arm and a smaller junction (inductance $L_J/\alpha$) in the other arm (see Fig.~\ref{fig1}c). The loop is threaded with an applied magnetic flux $\Phi$. The internal capacitances of the junctions are small and the self-resonance frequencies of the element are expected to be above 30 GHz. Therefore, at typical frequencies of circuit QED experiments, it behaves as a nonlinear inductor and its lowest energy configuration corresponds to having equal phase drops across the three large junctions. Thus, the SNAIL provides a $6\pi$-periodic potential energy
\begin{equation}
U_{{\rm SNAIL}}(\varphi_s)=-\alpha E_{J}\cos\varphi_s-3E_{J}\cos\frac{\varphi_{{\rm ext}}-\varphi_s}{3}\label{eq:snail_potential},
\end{equation}
where $E_J$ is the Josephson energy, and $\varphi_{\rm ext}=2\pi\Phi/\Phi_0$. We operate in a regime where phase fluctuations are suppressed, and the potential can be Taylor expanded near one of its equivalent minima. We denote the expansion coefficients by $c_{n}=\frac{1}{E_{J}}\frac{d^nU_{{\rm SNAIL}}}{d\varphi_s^n}(\varphi_{\min})$ and the minimum location $\varphi_{{\rm min}}$ is determined from the condition  
\begin{equation}
c_{1}\equiv\alpha \sin\varphi_{{\rm min}}+\sin\frac{\varphi_{{\rm min}}-\varphi_{{\rm ext}}}{3}=0.
\end{equation}
These flux-dependent coefficients $c_{n}=c_{n}(\varphi_{{\rm ext}})$,  together with $L_J$, completely characterize the behavior of the SNAIL when embedded in a larger electrical circuit. For instance, the linear inductance of the SNAIL is $L_{s}(\varphi_{{\rm ext}})=L_{J}/c_{2}(\varphi_{{\rm ext}})$.

\begin{figure}
\includegraphics[angle = 0, width = \figwidth]{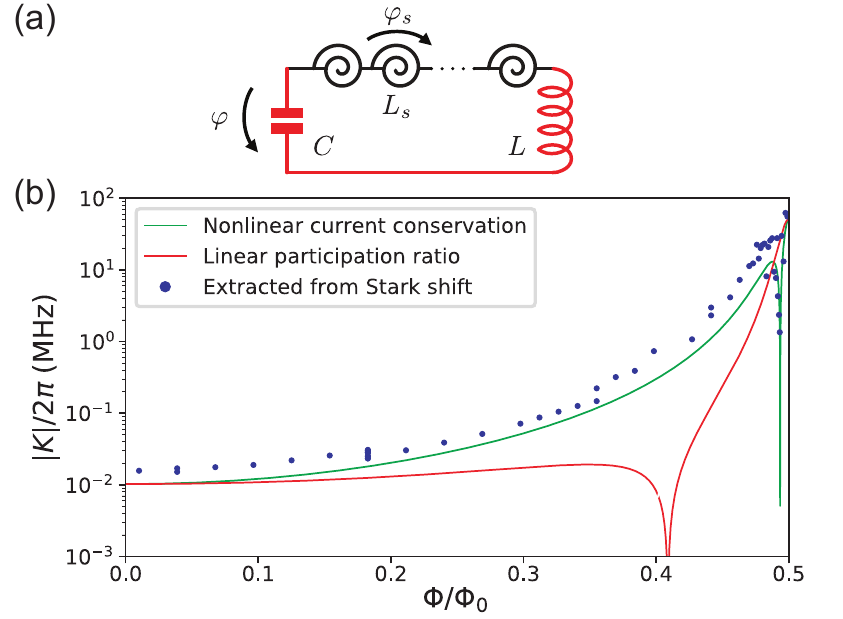}
\caption{\label{figS1_g4}  (a) Single-mode lumped-element circuit model for an SPA with $M$ SNAILs, where the microstrip resonator is approximated by a series $LC$ circuit (red). Phase drop across each SNAIL is denoted as $\varphi_s$, while $\varphi$ is the canonical phase coordinate for the mode. (b) Magnitude of Stark shift per photon $|K|$ of device A from Fig.~\ref{fig4_g4}, together with theoretical predictions: green includes nonlinear current conservation, red is the linear participation ratio-based approach in which $K \propto c_4$. } 
\end{figure}

To realize an SPA with a resonance in the 4-10 GHz range, an array of $M$ identical SNAILs is embedded into a transmission line resonator. The simplified circuit model, shown in Fig.~\ref{figS1_g4}a, consists of a series combination of capacitance $C$, inductance $L$, and an array of $M$ SNAILs. We will further assume that the SNAIL array can be considered as a lumped subcircuit, in which the phase splits equally among the individual SNAILs. This assumption is justified for small capacitance to ground of the inter-SNAIL islands \cite{masluk_microwave_2012} and for identical SNAILs.

In this case the total inductance of the emergent electromagnetic mode becomes flux tunable, as it consists of flux-independent inductance $L$ and flux-dependent inductance $L_{s}^{{\rm array}}(\varphi_{{\rm ext}})=ML_{s}(\varphi_{{\rm ext}})$ coming from the SNAIL array. The resonance frequency of this SPA mode is
\begin{align}
\omega_{a}(\varphi_{{\rm ext}})=\frac{1}{\sqrt{C\left[L+L_{s}^{{\rm array}}(\varphi_{{\rm ext}})\right]}}\nonumber\\
=\frac{\omega_{0}}{\sqrt{1+M\xi_{J}/c_{2}(\varphi_{{\rm ext}})}},
\end{align}
where we have defined the dimensionless coefficient $\xi_{J}=L_{J}/L$
and the resonance frequency in the absence of the array $\omega_{0}=1/\sqrt{LC}$.

The Lagrangian of this system is
\begin{subequations}
\begin{align}
{\cal L} &= \frac{C\varphi_0^2}{2}\dot{\varphi}^2-U(\varphi_s,\varphi)\label{lagrangian_one_mode},  \\
U(\varphi_{s},\varphi)&=MU_{{\rm SNAIL}}(\varphi_{s})+\frac{1}{2}E_{L}(\varphi-M\varphi_{s})^{2},
\end{align}
\end{subequations}
where $\varphi$ is the mode canonical phase coordinate, $\varphi_0$ is the reduced flux quantum, and $E_L=\varphi_0^2/L$.

The coordinate $\varphi_{s}$ is not an independent variable, as it does not have its own kinetic energy. Therefore, prior to quantization, we need to eliminate it by minimizing the nonlinear potential energy $U(\varphi_{s},\varphi)$ as a function of $\varphi_s$. The resulting trivial Lagrange equation of motion is equivalent to imposing a full nonlinear current conservation condition at the node between the SNAIL array and the linear inductor. Using Eq.~(\ref{eq:snail_potential}), we can write it as
\begin{equation}
\alpha\sin\varphi_{s}+\sin\frac{\varphi_{s}-\varphi_{{\rm ext}}}{3}+\xi_{J}(M\varphi_{s}-\varphi)=0. \label{eq:current conservation}
\end{equation}

This equation determines the SNAIL phase $\varphi_{s}[\varphi]$ as a function of mode canonical coordinate $\varphi$, which has to be further quantized. The potential energy can now be written in terms of a single degree of freedom $\varphi$ 
\begin{equation}
U(\varphi)=MU_{{\rm SNAIL}}(\varphi_{s}[\varphi])+\frac{1}{2}E_{L}(\varphi-M\varphi_{s}[\varphi])^{2}. \label{pot}
\end{equation}

Given that we operate in the regime of small phase fluctuations, we can again Taylor expand the renormalized potential $U(\varphi)$, resulting in the coefficients $\tilde{c}_{n}=\frac{1}{E_{J}}\frac{d^nU}{d\varphi^n}(\overline{\varphi}_{\min})$,
where $\overline{\varphi}_{{\rm min}}$ is determined from the condition
$\tilde{c}_{1}=0$. Using the current conservation equation \eqref{eq:current conservation}, we can write the first four Taylor coefficients as
\begin{align}
\tilde{c}_{1}&=\xi_{J}(\overline{\varphi}_{\rm min}-M\varphi_{s}[\overline{\varphi}_{\rm min}]),\\ 
\tilde{c}_{2}&=\xi_{J}(1-M\frac{\partial\varphi_{s}}{\partial\varphi}[\overline{\varphi}_{\rm min}]),\\
\tilde{c}_{3}&=-M\xi_{J}\frac{\partial^{2}\varphi_{s}}{\partial^{2}\varphi}[\overline{\varphi}_{\rm min}],\\ 
\tilde{c}_{4}&=-M\xi_{J}\frac{\partial^{3}\varphi_{s}}{\partial^{3}\varphi}[\overline{\varphi}_{\rm min}].
\end{align}

The derivatives of the implicit function $\varphi_s[\varphi]$ can be found by differentiating Eq.~\eqref{eq:current conservation}. For example, by differentiating it once we obtain
\begin{align}
\frac{\partial\varphi_{s}}{\partial\varphi}=\frac{\xi_{J}}{\alpha\cos\varphi_{s}+\frac{1}{3}\cos\frac{\varphi_{s}-\varphi_{{\rm ext}}}{3}+M\xi_{J}}.\label{eq:intermediate}
\end{align} 

In addition, we can show from Eq.~\eqref{eq:current conservation} that the presence of the series linear inductor does not change the location of the SNAIL potential minimum $\varphi_{s}[\overline{\varphi}_{{\rm min}}]=\varphi_{{\rm min}}$. Therefore, we can express the Taylor coefficients $\tilde{c}_{n}$ for the renormalized potential $U(\varphi)$ in terms of the bare ones $c_n$ introduced earlier for a single SNAIL
\begin{align}
\tilde{c}_{2}&=\frac{p}{M}c_{2},\\ 
\tilde{c}_{3}&=\frac{p^{3}}{M^{2}}c_{3},\\ 
\tilde{c}_{4}&=\frac{p^{4}}{M^{3}}\bigg(c_{4}-\frac{3c_{3}^{2}}{c_{2}}(1-p)\bigg),\label{eq:c4_effective}
\end{align}
where we have defined the array linear participation ratio:
\begin{equation}
p\equiv\frac{ML_s}{L + ML_s}=\frac{M\xi_{J}}{c_{2}+M\xi_{J}}.
\end{equation}

After performing the Legendre transformation and canonical quantization, the Hamiltonian of the mode can be written as
\begin{equation}
\bm H=4E_{C}{\bm N}^{2}+E_{J}\bigg(\frac{\tilde{c}_{2}}{2!}{\bm \varphi}^{2}+\frac{\tilde{c}_{3}}{3!}{\bm \varphi}^{3}+\frac{\tilde{c}_{4}}{4!}{\bm \varphi}^{4}+...\bigg),
\end{equation}
where $E_{C}=e^{2}/2C$ and $[\bm \varphi,\bm N]=i$. For more convenience, we can introduce the bosonic raising and lowering operators $\bm a^{\dagger}$ and $\bm a$ that diagonalize the quadratic part of the Hamiltonian in the excitation number basis. The SPA Hamiltonian after this second quantization can be written, truncated to fourth order, as
\begin{equation}
{\bm H_{\rm SPA}}/\hbar=\omega_{a}{\bm a}^{\dagger}{\bm a}+g_{3}(\bm a+\bm a^{\dagger})^{3}+g_{4}(\bm a+\bm a^{\dagger})^{4},
\label{ham_appendix}
\end{equation}
where
\begin{align}
\hbar g_{3} &=\frac{1}{6}\frac{p^{2}}{M}\frac{c_{3}}{c_{2}}\sqrt{E_{C}\hbar\omega_{a}}\label{g3_app}, \\
\hbar g_{4} &=\frac{1}{12}\frac{p^{3}}{M^{2}}\bigg(c_{4}-\frac{3c_{3}^{2}}{c_2}(1-p)\bigg)\frac{1}{c_{2}}E_{C}. \label{g4_app}
\end{align}

In general, this Hamiltonian has both odd and even nonlinearities for $\Phi/\Phi_0\neq n/2$ (where $n$ is any integer),  unlike the symmetric transmon Hamiltonian \cite{koch_charge-insensitive_2007}. These nonlinearities inherit their flux dependence from that of the SNAIL potential, and thus are tunable \textit{in situ}.

We can now relate these nonlinearities to the Stark shift per photon $K$ that is measured in the experiment. In a nearly harmonic oscillator, $K$ can be calculated as the dispersion of transition frequencies between the neighboring energy levels,
\begin{align}
\hbar K = \frac{d^2E(n)}{dn^2},
\end{align}
where $E(n)$ is the $n$th energy level of the Hamiltonian \eqref{ham_appendix}. In the case $g_3=0$,
which is the well-known Duffing oscillator model, $K$ is simply related to the Hamiltonian parameter $g_4$ via $K=12g_4$. However, in the asymmetric SPA potential this relation is modified to $K=12\left(g_4-5g_3^2/\omega_a\right)$, where the last term comes from the second-order perturbation theory correction to the energy levels. Using expressions (\ref{g3_app}-\ref{g4_app}) we thus obtain
\begin{align}
\hbar K=\frac{p^{3}}{M^{2}}\bigg(c_{4}-\frac{3c_{3}^{2}}{c_2}(1-p)-\frac{5}{3}\frac{c_3^2}{c_2}p  \bigg)\frac{1}{c_{2}}E_{C}. \label{Kerr_app}
\end{align}

We would like to stress that the calculation of the Kerr effect that we have outlined is significantly different from previous calculations in 3-wave-mixing amplifiers, such as for the JPC, in Ref.~\onlinecite{schackert_practical_2013, flurin_josephson_2015, liu_josephson_2017}. Previous calculations considered first the linearized circuit, in which the total phase drop $\varphi$ splits between the nonlinear circuit elements and the linear inductor in proportion to their respective participation ratios $p$ and $1-p$. Then they assumed that the nonlinearity is diluted by the corresponding power of the participation ratio, in which case we would have $g_{n}\propto p^{n-1}c_{n}$. As we can see from \eqref{g3_app} and \eqref{g4_app}, this approach yields the correct values for the lowest order cubic nonlinearity, but fails to predict higher-order nonlinearities, such as $g_4$ and therefore $K$, correctly.

This discrepancy arises because the linear participation ratio-based approach does not properly account for the nonlinear current conservation, Eq.~\eqref{eq:current conservation}, in the SPA between the SNAIL array and the inductor. This effect leads to a renormalization of $g_{4}$ due to $c_{3}$ evident in \eqref{g4_app}. One can see that in the limit of small participation ratio $p\to0$ the additional contribution is equal to $-3c_{3}^{2}/c_{2}$, which does not contain any small parameters relative to $c_{4}$. This significantly shifts the Kerr-free point in flux and modifies the whole Kerr nonlinearity profile. The comparison of data taken on device A with predictions of both approaches is shown in Fig.~\ref{figS1_g4}b. Note that at certain fluxes the predictions differ by several orders of magnitude.

Moreover, apart from this renormalization effect, there is a trivial second-order perturbation-theory correction to energy levels due to the $g_{3}$ term in the Hamiltonian, which affects the Stark shift per photon $K$ (last term in Eq.~\ref{Kerr_app}). Note that both $g_4$ and this correction scale identically with $E_C$. This perturbative correction is insignificant in the limit of small participation ratio $p$. On the other hand, in the limit $p\to 1$, the renormalization effect due to the linear inductance becomes irrelevant and the perturbative contribution becomes important instead. 

\begin{figure}
\includegraphics[angle = 0, width = \figwidth]{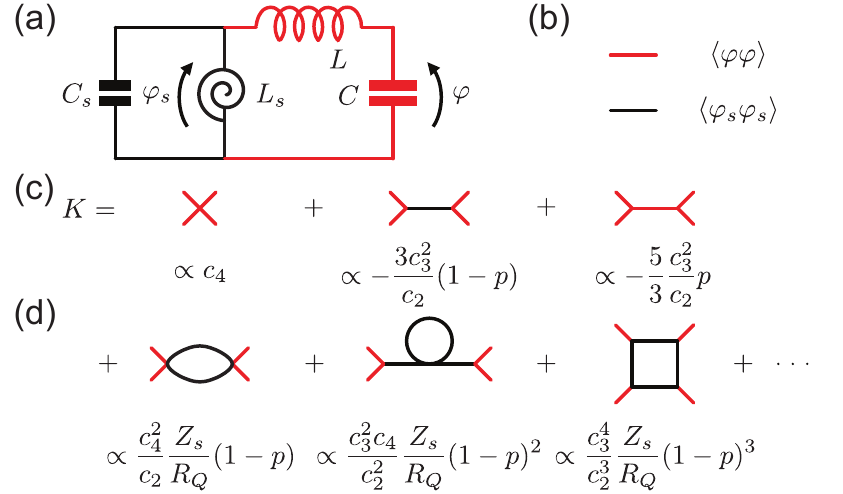}
\caption{\label{figS1_diag}  (a) Two-mode lumped-element circuit model for an SPA with $M=1$ SNAIL. One mode is the SPA mode, frequency $\omega_a \approx 1/\sqrt{C(L+L_s)}$, and the other is a high-frequency SNAIL mode, frequency $\Omega_s \approx 1/\sqrt{C_s L_s} \gg \omega_a$.
(b) Definitions for elements of Feynman diagrams used to eliminate the high-frequency SNAIL mode. Red (black) lines represent propagator for low- (high-) frequency mode with canonical coordinate $\varphi$ ($\varphi_s$).
(c) Diagrammatic series to calculate Stark shift per photon $K$ of SPA (red) mode. Each diagram corresponds to a term in Eq.~\ref{Kerr_app}, where third- (fourth-) order interaction vertices are weighted by $c_3$ ($c_4$) from the SNAIL potential.
(d) Example higher-order one-loop diagrams, each suppressed by $Z_s/R_Q$ where $Z_s = \sqrt{L_s/C_s}$ and $R_Q = \hbar/4e^2$.
}
\end{figure}
\subsection{Renormalization of Kerr \label{renormalization_K}}
We can also explain the physics of the nontrivial contribution to $g_4$ in Eq.~\eqref{g4_app} from a different perspective. This correction is nothing but a renormalization of the potential due to high-frequency modes that are inevitably present in the system (for example, the plasma modes of the SNAIL junctions). Such modes cannot be directly probed by low-energy experiments, but their effect is evident in measurable quantities such as the Stark shift per photon $K$.

To demonstrate this point, consider the circuit in Fig.~\ref{figS1_diag}a, which includes the capacitance $C_s$ shunting the SNAIL (we consider $M=1$ for simplicity). Such a circuit has two eigenmodes: the SPA low-frequency mode ($\omega_a/2\pi\sim 7\rm \, GHz$ ) and the high-frequency SNAIL mode ($\Omega_s/2\pi>30\,\rm GHz$). In the low-participation ratio limit, the SPA mode is  mostly localized in the series $LC$ circuit (red in Fig.~\ref{figS1_diag}a), with the high-frequency mode in the SNAIL and its shunting capacitor (black).

In this system, in contrast with Eq.~\eqref{lagrangian_one_mode} (see also Fig.~\ref{figS1_g4}a), the variable $\varphi_{s}$ becomes a real quantum-mechanical coordinate with its own conjugate momentum. Therefore, it should be quantized on equal footing with $\varphi$.  However, since $\Omega_s\gg\omega_a$, this fast degree of freedom can be integrated out using the Born-Oppenheimer method or more sophisticated QFT techniques \cite{altland_condensed_2010}.

In the Feynman diagram language, the Kerr nonlinearity of the SPA mode can be represented as a fourth-order self-interaction vertex. The example diagrams that contribute to the renormalization of this vertex are depicted in Figs.~\ref{figS1_diag}c and \ref{figS1_diag}d.

The first diagram in Fig.~\ref{figS1_diag}c comes directly from the quartic term in the potential energy of the SNAIL element \eqref{eq:snail_potential} and is therefore proportional to the corresponding Taylor coefficient $c_4$. Including only this diagram results in $K\propto c_4$, which is equivalent to the linear participation ratio-based approach discussed previously. 

The interesting and nontrivial correction to $K$ comes from the high-energy SNAIL mode, and is represented by the second diagram in Fig.~\ref{figS1_diag}c. This contribution is equivalent to the second term in \eqref{Kerr_app}, and it does not depend on $\Omega_s$ as long as the requirement $\Omega_s\gg\omega_a$ is satisfied. In the previous single-mode model, this contribution arose from imposing the full nonlinear current conservation \eqref{eq:current conservation}.

The last diagram in Fig.~\ref{figS1_diag}c depicts two cascaded $c_3$ self-interactions of the SPA mode. This is equivalent to the trivial second-order perturbation theory correction, as seen in the last term of Eq.~\eqref{Kerr_app}. Together, the three diagrams in Fig.~\ref{figS1_diag}c give Kerr to the leading order and fully replicate the result Eq.~\eqref{Kerr_app}. 

In principle, there are other contributions to the renormalization of the fourth-order self-interaction vertex. Some example one-loop processes are shown in Fig.~\ref{figS1_diag}d. However, such diagrammatic corrections are suppressed by a factor $(Z_{s}/R_{Q})^l$, where $Z_{s}$ is the impedance of the fast mode and $l$ is the number of loops. In practice, $Z_s/R_Q \ll 1$ in our fabrication process. In fact, it is fundamentally difficult to achieve $Z_s \gtrsim R_Q$ \cite{masluk_microwave_2012}.

Comparing the two electrical circuits in Figs.~\ref{figS1_g4}a and \ref{figS1_diag}a, we see that the circuit in Fig.~\ref{figS1_g4}a corresponds to taking the limit $C_s\to0$ for the one in Fig.~\ref{figS1_diag}a. Rigorously taking this limit would lead to large quantum fluctuations of phase, and the diagrammatic series would diverge. To ensure the convergence, $C_s$ still has to be large enough to satisfy $Z_s\ll R_Q$. In this case, classical elimination of the fast degree of freedom $\varphi_s$ via nonlinear current conservation \eqref{eq:current conservation} is justified.

Understanding these effects is important because the general approach for designing circuit QED systems relies on pushing the spurious modes up in frequency and then neglecting their influence by arguing that the detuning to these modes is large. We have shown that the presence of these modes can influence low-energy observables, and they have to be accounted for either by means of full nonlinear current conservation, or equivalently by integrating out the high-energy modes.

\subsection{Distributed-element model of SPA\label{distributed_model}}
While the lumped-element model above elucidates the important details in the theoretical treatment of the SPA and provides physical intuition, it cannot strictly be applied to our devices. In the SPA, the SNAIL array is embedded in a transmission line resonator, which is a distributed circuit element, see Fig.~\ref{fig1}d.

The Lagrangian of such a system can be written as
\begin{align}
{\cal L} =& \left(\int_{-l_{\textrm{MS}}/2}^{-0}+\int_{+0}^{l_{\textrm{MS}}/2}\right)\left[ \frac{c}{2}(\partial_t\phi)-\frac{1}{2\ell}(\partial_x\phi)^2 \right]dx\nonumber \\
&-MU_{\rm SNAIL} \left( \frac{\varphi_r-\varphi_l}{M} \right),
\end{align}
where $c$ is the capacitance per unit length and $\ell$ is the inductance per unit length. The generalized flux $\phi(x,t)$ on the transmission line is a one-dimensional massless Klein-Gordon field which has a discontinuity at $x=0$, where the transmission line is interrupted by the lumped-element SNAIL array. For convenience, we have introduced $\varphi_l=\phi(-0,t) / \varphi_0$ and $\varphi_r=\phi(0+,t)/\varphi_0$ to denote the superconducting  phase on both sides of the array. 

Using zero current boundary conditions at $x = \pm l_{\rm MS}/2$ and linearizing the Lagrange equation of motion for this system, we can perform an eigenmode decomposition and find the resonant frequency $\omega_a$ of the structure as the smallest nontrivial solution of the equation
\begin{equation}
\omega_a\tan\left(\frac{\pi}{2}\frac{\omega_a}{\omega_0}\right)=\frac{2Z_c}{ML_s(\varphi_{\rm ext})}, \label{eq:freq_app}
\end{equation}
where $Z_c=\sqrt{\ell/c}$ is the characteristic impedance of the transmission line, and $\omega_0=\pi/l_{\rm MS}\sqrt{\ell c}$ is the resonant frequency when the array of SNAILs is replaced with a short. We use \eqref{eq:freq_app} for the fits of $\omega_a$ in Fig.~\ref{fig2}.

Following Refs. \onlinecite{wallquist_selective_2006,nigg_black-box_2012}, we calculate the nonlinearities of the SPA as
\begin{align}
g_3=&\frac{4Z_c c_3}{3M^2L_J}\sqrt{\frac{Z_c}{R_Q}}
\left[\frac{\cos^2 \left(\frac{\pi}{2}\frac{\omega_a}{\omega_0}\right) }{\pi\frac{\omega_a}{\omega_0} + \sin\left(\pi\frac{\omega_a}{\omega_0}\right) }\right]^{3/2} \label{eq:g3_dist},\\
K=& \frac{\omega_a \sin^2\left(\pi\frac{\omega_a}{\omega_0}\right) \cot\left(\frac{\pi}{2}\frac{\omega_a}{\omega_0}\right) }{c_2M^2 \left[ \pi\frac{\omega_a}{\omega_0}+\sin\left(\pi\frac{\omega_a}{\omega_0}\right) \right]^2} \frac{Z_c}{R_Q}
\nonumber\\ & \times
\left[c_4-\frac{c_3^2}{c_2}\frac{3+5\left(\frac{\omega_a ML_s}{2Z_c}\right)^2}{1+3\left(\frac{\omega_a ML_s}{2Z_c}\right)^2} \right], \label{Kerr_distributed}
\end{align}
where $R_Q=\hbar/(2e)^2$ is the resistance quantum. The limits of small and unity participation ratio correspond in this model to $ML_s\ll Z_c/\omega_a$ and $ML_s\gg Z_c/\omega_a$, respectively. In these limits, the corrections to Kerr due to $c_3$ coincide in both the distributed- \eqref{Kerr_distributed} and lumped-element \eqref{Kerr_app} models.

We use the expressions \eqref{eq:g3_dist} and \eqref{Kerr_distributed} for $g_3$ and $K$ to plot the first-principles theory curves in Figs.~\ref{fig3}c and \ref{fig4_g4}b, respectively.
\subsection{Scaling with number of SNAILs\label{scalingM}}
The above expressions for $g_3$ and $K$ give these nonlinearities as functions of designable parameters, such as the number of SNAILs $M$. However, the scaling with $M$ can be rather unintuitive when applied to practical device design changes. For example, we consider the goal of decreasing $K$ by increasing $M$. Using the lumped-element expression \eqref{Kerr_app} for simplicity, the dependence on $M$ is $K \propto 1/M^2$ when $p$, $\alpha$, and $E_C$ are held constant. To then hold the operating frequency $\omega_a$ constant, increasing $M$ requires a corresponding increase in $E_J$ (smaller $L_J$), as is suggested in Ref.~\onlinecite{eichler_controlling_2014} for higher dynamic range in JPAs. Unfortunately, parasitic geometric inductance in the leads used to make the junctions prevents the reduction of $L_J$ much below its current $\approx 40$ pH value without significantly affecting $p$.

In practice, when increasing $M$, we hold $L_J$ constant and increase $E_C$ to realize the desired frequency $\omega_a$. The predicted scaling in this instance is weaker ($K \propto p^2/M$). Moreover, this tactic also tends to increase $p$, which further weakens this scaling. As such, we find that increasing $M$ much past 20 for our design does not strongly reduce $K$ until $M \gtrsim 200$, at which point the SNAIL array can no longer be considered a lumped element. Future work must model these larger arrays as SNAIL transmission lines, where increasing $M$ reduces $K$ if the characteristic impedance is independent of $M$.

\section{Dynamic Range of an Amplifier} \label{sec:app_DR}
In this Appendix we discuss the map between the Hamiltonian and input-output port coupling parameters and the amplifier characteristics, specifically the saturation power and intermodulation distortion. In Sec.~\ref{Appendix_harmonic_balance_1}, we theoretically treat amplifier saturation by performing harmonic-balance analysis of arbitrary-signal-power scattering. Section \ref{sec:app_IMD} contains theoretical treatment of IMD. Section \ref{comparison_to_data} compares in detail the measured saturation power and theory. 
\subsection{Semiclassical solution: harmonic balance \label{Appendix_harmonic_balance_1}}
We analyze the response of the system using standard input-output theory together with the quantum Langevin equation (QLE) for mode $\bm a$,
\begin{align}
\dot{\bm a} & =\frac{i}{\hbar}[\bm H_{\rm SPA} +\bm{H_{\rm drive}},\bm a]-\frac{\kappa}{2}(\bm a-\bm a^\dagger),
\label{qle1}\\
\bm{H_{\rm drive}} &= \hbar (\bm u_{\rm in}+\bm u_{\rm in}^{\dagger})(\bm a+\bm a^{\dagger}),
\label{qle2}\\
\bm u_{{\rm out}} &= i\kappa(\bm a-\bm a^\dagger)-\bm u_{{\rm in}},
\label{qle3}
\end{align}
written here without the rotating wave approximation (RWA) since we are applying an off-resonant pump. The classical drive amplitude $u_{\rm in( out)}= \sum_\omega u^\omega_{\rm in(out)}e^{-i\omega t}$ is related to the input (output) power at the corresponding frequency $\omega$ via $P_{{\rm in(out)}} = \frac{\hbar\omega_a}{\kappa} |u^{\omega}_{\rm in(out)}|^{2}\cdot \frac{\omega_{a}^{2}}{\omega^{2}}$ (for the capacitive coupling).

A linear harmonic oscillator that is pumped at $\omega_{p}$ and probed at $\omega_{s}$ responds independently to each of these frequency components of the incoming field. In contrast the SPA, which consists of a weakly nonlinear resonator, will produce a response at all harmonics $\omega_{mn}=m\omega_{p}+n\omega_{s}$. Conventional degenerate parametric-amplifier theory \cite{clerk_introduction_2010,roy_introduction_2016} takes into account only one additional harmonic (the idler)  at $\omega_{i}=\omega_{p}-\omega_{s}$, although in practice all higher intermodulation products (IMDs) will be created. Their magnitudes are often small and therefore neglecting these harmonics is a reasonable starting point. 

Restricting ourselves to signal $\omega_{s}$, idler $\omega_{i}$, and pump $\omega_{p}$ frequencies, we can solve the QLE (\ref{qle1}-\ref{qle3}) using the semiclassical harmonic-balance method \cite{kamal_signal--pump_2009,sundqvist_negative-resistance_2014,liu_josephson_2017}. We will denote the input drive strengths at $\omega_s$, $\omega_i$, and $\omega_p$ as $u_s$, $u_i$, and $u_p$, respectively, and treat them as classical drives (i.e.~c-numbers). After applying harmonic-balance conditions, we obtain a self-consistent system of coupled equations that relates all harmonics to each other:
\begin{widetext}
\begin{align}
\bigg(\omega_{p}-\omega_{a}+i\frac{2}{3}\kappa-g_{4}\big[\frac{32}{9}|\alpha_{p}|^{2}+16|\alpha_{s}|^{2}+16|\alpha_{i}|^{2}\big]\bigg)\alpha_{p}&=u_{p}+6g_{3}\alpha_{i}\alpha_{s},\label{pump_amp} \\
\bigg(\omega_{s}-\omega_{a}+i\frac{\kappa}{2}-g_{4}\big[\frac{32}{3}|\alpha_{p}|^{2}+12|\alpha_{s}|^{2}+12|\alpha_{i}|^{2}\big]\bigg)\alpha_{s}&=u_{s}+4g_{3}\alpha_{p}\alpha_{i}^{*},\label{signal_amp} \\
\bigg(\omega_{i}-\omega_{a}+i\frac{\kappa}{2}-g_{4}\big[\frac{32}{3}|\alpha_{p}|^{2}+12|\alpha_{i}|^{2}+12|\alpha_{s}|^{2}\big]\bigg)\alpha_{i}&=u_{i}+4g_{3}\alpha_{p}\alpha_{s}^{*},\label{idler_amp}
\end{align}
\end{widetext}
where $\alpha_s$, $\alpha_i$ and $\alpha_p$ denote the mean intracavity amplitudes at the corresponding frequencies. 

To cast these equations into a more familiar form, we introduce the signal detuning $\omega=\omega_{s}-\frac{\omega_{p}}{2}$, the pump detuning $\Delta=\omega_{a}-\frac{\omega_{p}}{2}$, and the effective pump detuning that includes Stark shift effects
\begin{equation}
\Delta_{{\rm eff}}=\Delta+g_{4}\big[\frac{32}{3}|\alpha_{p}|^{2}+12|\alpha_{s}|^{2}+12|\alpha_{i}|^{2}\big].\label{effective_detuning}
\end{equation}

With the help of these notations, Eqs.~(\ref{signal_amp})-(\ref{idler_amp}) can be written via a susceptibility matrix as
\begin{align}
\left(\begin{array}{c}
\alpha_{s}\\
\alpha_{i}^{*}
\end{array}\right)&=\frac{1}{{\cal D}}\left(\begin{array}{cc}
-\omega-\Delta_{{\rm eff}}-i\frac{\kappa}{2} & 2g_{{\rm eff}}\\
2g_{{\rm eff}}^{*} & \omega-\Delta_{{\rm eff}}+i\frac{\kappa}{2}
\end{array}\right)\left(\begin{array}{c}
u_{s}\\
u_{i}^{*}
\end{array}\right),\nonumber \\  
{\cal D}&=\Delta_{{\rm eff}}^{2}-\omega^{2}+\frac{\kappa^{2}}{4}-4|g_{{\rm eff}}|^{2}-i\kappa\omega,\label{eq:sss}
\end{align}
where $g_{{\rm eff}}=2g_{3}\alpha_{p}$. Using Eq.~\eqref{eq:sss} and the input-output relation \eqref{qle3}, we calculate the phase-preserving power gain $G=|u^{\omega_s}_{{\rm out}}|^{2}/|u_{s}|^{2}$. After some algebraic transformations, it can be cast into the form
\begin{equation}
G=1+\frac{4\kappa^{2}|g_{{\rm eff}}|^{2}}{(\Delta_{{\rm eff}}^{2}-\omega^{2}+\frac{\kappa^{2}}{4}-4|g_{{\rm eff}}|^{2})^{2}+(\kappa\omega)^{2}}\label{eq:gain semiclassics}.
\end{equation}

We note the close similarity between this expression for gain and Eq.~\eqref{gain_simple}. However, here all parameters are effective ones that depend on the input signal power. For example, the parameter $g$ in Eq.~\eqref{gain_simple} depends on $\alpha_{p}$ which is treated as a constant. In Eq.~(\ref{eq:gain semiclassics}) however, $\alpha_p$ is determined from the self-consistent system of Eqs.~(\ref{pump_amp})-(\ref{idler_amp}), and therefore depends on the signal power. 

To achieve large gain, the denominator in Eq.~(\ref{eq:gain semiclassics}) should be tuned close to zero, often called the parametric instability point. Altering this denominator by even a small amount in response to increased input signal power will reduce the gain and cause amplifier saturation.

The small change in gain can therefore be written as 
\begin{equation}
\delta G=G_0^{3/2}\frac{8g_{3}}{\kappa\sqrt{n_{p}}}\delta n_{p}-G_0^{3/2}\frac{1}{2\kappa g_{3}\sqrt{n_{p}}}\delta\Delta_{{\rm eff}}^{2},\label{eq:delta gain}
\end{equation}
where $G_0$ denotes the small-signal gain and $n_p=|\alpha_p|^2$. We have also restricted to the case $\omega=0$ and neglected the weak influence of numerator in Eq.~\eqref{eq:gain semiclassics}.

The first term in Eq.~\eqref{eq:delta gain} shows the effect of changes in pump population $n_p$. Pump population can be depleted in response to increasing signal power due to the intrinsic $g_3$-induced coupling of $\alpha_p$ and $\alpha_s$ in Eq.~(\ref{pump_amp}). This back-action on the pump due to the signal \cite{kamal_signal--pump_2009}, often called pump depletion, will cause amplifier saturation even when $g_4=0$. Moreover, pump depletion can also arise due to the amplification of quantum fluctuations within the amplifier band \cite{roy_quantum-limited_2018}.

The second term in Eq.~\eqref{eq:delta gain} shows the effect of changes in $\Delta_{\rm eff}$. This change originates from the Stark shift contributions of the input signal in the detuning \eqref{effective_detuning}. In the first-order harmonic-balance theory discussed so far, this effect vanishes when $g_{4}=0$. 

From Eq.~\eqref{eq:delta gain}, it is straightforward to estimate how the saturation power scales with the parameters of the SPA for these two mechanisms
\begin{align}
P_{-1\text{dB}}^{\text{Stark}}&\sim \frac{\kappa}{|g_{4}|} \frac{1}{G_0^{5/4}} \hbar \omega_a \kappa, \label{Stark_lim_app}\\
P_{-1\text{dB}}^{\rm{pump \,dep}}&\sim \frac{\kappa}{g_{3}^{2} / \omega_a}\frac{1}{G_0^{3/2}} \hbar \omega_a \kappa, \label{pumpDep_lim_app}
\end{align}
where for brevity we have assumed $\Delta=0$. 

Before comparing this semiclassical harmonic-balance theory to the experimental data, let us cover one more topic which is relevant for amplifiers: spurious intermodulation distortion.
\begin{figure}
\includegraphics[angle = 0, width = \figwidth]{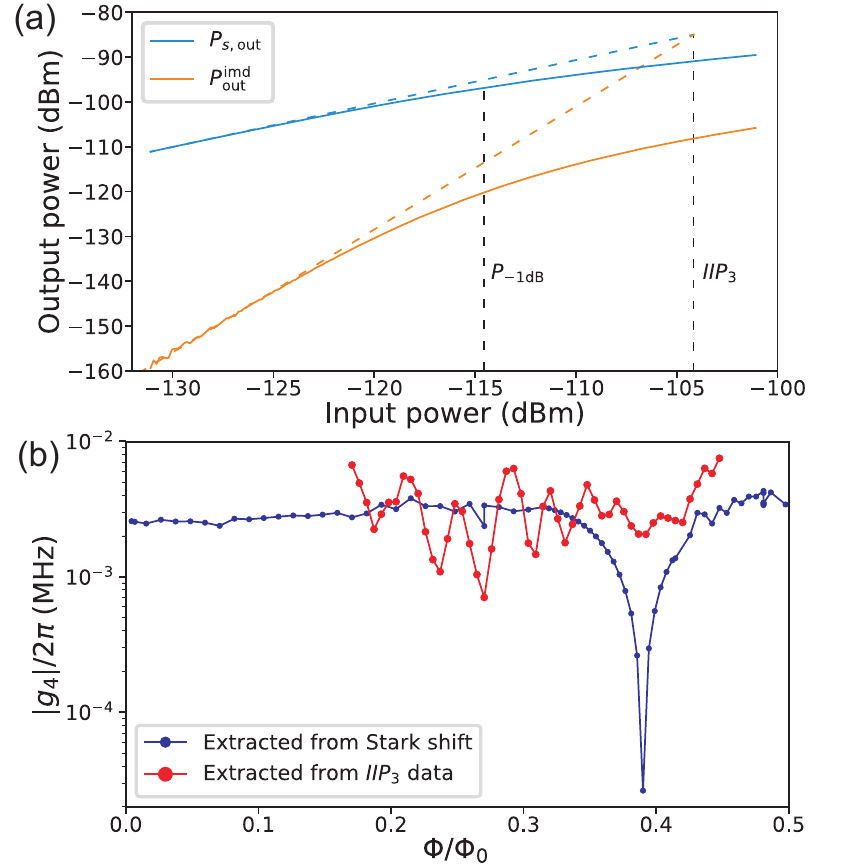}
\caption{\label{figS2_g4imd}  
(a) Example $IIP_3$ experiment showing measured signal power $P_{s, \rm out}$ (solid blue) and third-order IMD sideband power $P^{\rm imd}_{{\rm out}}$ (solid orange) as a function of applied input power $P_{s,\rm in}$. Intersection of low-power asymptotes (dashed) gives $IIP_3$. The 1-dB compression power $P_{\rm -1dB}$ is also indicated for reference. 
(b) 4-wave-mixing nonlinearity $g_4$ as a function of applied magnetic flux $\Phi$ for device D, extracted from the Stark shift (blue) or $IIP_3$ (red) experiments.
}
\end{figure}
\subsection{Intermodulation distortion (IMD)} \label{sec:app_IMD}
Let us now extend the analysis to consider the response of an amplifier to two monochromatic input tones. Standard parametric-amplifier theory predicts that the output will consist of two amplified tones at the same frequencies and two idler tones. However, because of the resonator nonlinearity, higher-order-mixing products will inevitably be created. The third-order IMD products are most relevant to the amplifier quality, as they are created within the amplifier bandwidth \cite{pozar_microwave_2012}.

Third-order IMDs measure the nonlinear 4-wave-mixing scattering processes of a device, which can be understood from the following frequency conditions, see Fig.~\ref{fig6_IMD}a. If we send two signals at $\omega_{s1}$ and $\omega_{s2}$, the nearest third-order-mixing products will be created at sideband frequencies $2\omega_{s1}-\omega_{s2}$ and $2\omega_{s2}-\omega_{s1}$. Focusing on the IMD product at $\omega_{\rm imd} = 2\omega_{s1}-\omega_{s2}$, this 4-wave-mixing process corresponds to the annihilation of two photons at $\omega_{s1}$ to create a photon each at $\omega_{s2}$ and $\omega_{\rm imd}$. Thus, measuring the power in the sideband at $\omega_{\rm imd}$ informs us about the strength of spurious 4-wave-mixing processes.

We can quantitatively relate this sideband power to the $g_4$ of the device. Including $\omega_{\rm imd}$ in the harmonic-balance analysis of the QLE \eqref{qle1}, we find that the Kerr term in the Hamiltonian acts as an effective drive at the respective IMD frequencies. For example, for the IMD product at $\omega_{\rm imd} = 2\omega_{s1}-\omega_{s2}$, the effective drive strength is equal to
\begin{equation}
u^{\rm imd}_{{\rm in (eff)}}=12g_{4}\alpha_{s1}^{2}\alpha_{s2}^{*}.
\end{equation}
Therefore, we can pretend that there is an input at the IMD frequencies, and the amplifier treats it as any other input signal, namely creates an amplified output $P^{\rm imd}_{{\rm out}}=GP^{\rm imd}_{\rm {in(eff)}}$ and the corresponding idler. For near-resonant tones,
\begin{equation}
P^{\rm imd}_{{\rm out}}=G\frac{\hbar\omega_{a}}{\kappa}|u^{\rm imd}_{{\rm in (eff)}}|^{2}=G\frac{\hbar\omega_{a}}{\kappa}(12g_{4})^{2}n_{s1}^{2}n_{s2},
\end{equation}
where $n_{s1(2)}=|\alpha_{s1(2)}|^2$ is the average intraresonator population, which can be related  to the input and output powers by $n_{s1(2)} \hbar\omega_{a}\kappa = GP_{s1(2),\rm in} = P_{s1(2),{\rm out}}$ (for $G\gg1$).
Thus, we can relate the output power at $\omega_{\rm imd}$ to the output powers at $\omega_{s1}$ and $\omega_{s2}$ by
\begin{equation}
P^{\rm imd}_{{\rm out}}=\frac{|12g_{4}|^{2}G}{\kappa^{4}(\hbar \omega_{a})^{2}}P_{s1,{\rm out}}^{2}P_{s2,{\rm out}}. \label{eq:bbb}
\end{equation}

Assuming $P_{s2,{\rm in}}=P_{s1,{\rm in}} = P_{s,{\rm in}} $, we note that $P^{\rm imd}_{{\rm out}}$ scales cubically with applied input power, whereas $P_{s,{\rm out}} = G P_{s, \rm in}$ scales linearly (see Fig.~\ref{figS2_g4imd}a). Thus, as in Ref.~\onlinecite{pozar_microwave_2012}, we can define the input-referred third-order intercept point $IIP_{3}$ as the input power where the low-power asymptotes of $P_{s,{\rm out}}$ (dashed blue) and $P^{\rm imd}_{{\rm out}}$ (dashed orange) intersect.
Using Eq.~(\ref{eq:bbb}) we then obtain
\begin{equation}
IIP_{3}=\frac{\kappa}{12|g_4|}\frac{1}{G^{3/2}}\hbar\omega_{a}\kappa.
\label{IIP3_app}
\end{equation}

While the above derivation for brevity assumes $G\gg1$, it can be shown that for arbitrary gain $G$ the $IIP_3$ is given by
\begin{equation}
IIP_3 = \frac{\kappa}{12 |g_4|} \left(\frac{1}{\sqrt{G}+1}\right)^3  \hbar \omega_a \kappa .
\end{equation}
This result can be used to extract the fourth-order nonlinearity $g_{4}$ in the presence of the amplification pump and compare it to the Stark shift experiment described in the main text (Fig.~\ref{fig4_g4}).

As shown in Fig.~\ref{figS2_g4imd}b, there is reasonably good agreement between $g_4$ measured by these independent experiments at fluxes away from the Kerr-free point. Ripples in the $g_4$ extracted from $IIP_3$ arise from variation of the impedance seen by the SPA (see discussion of Fig.~\ref{fig5_P1dB} in main text). Strikingly, near the Kerr-free point the $g_4$ extracted from $IIP_3$ does not exhibit the reduction of the magnitude observed in the Stark shift measurement. We conjecture that this effect is related to the discrepancy between theory and experiment for $P_{\rm -1dB}$ seen in Fig.~\ref{fig5_P1dB}b. We explore this question further in the next section.
\begin{figure}
\includegraphics[angle = 0, width = \figwidth]{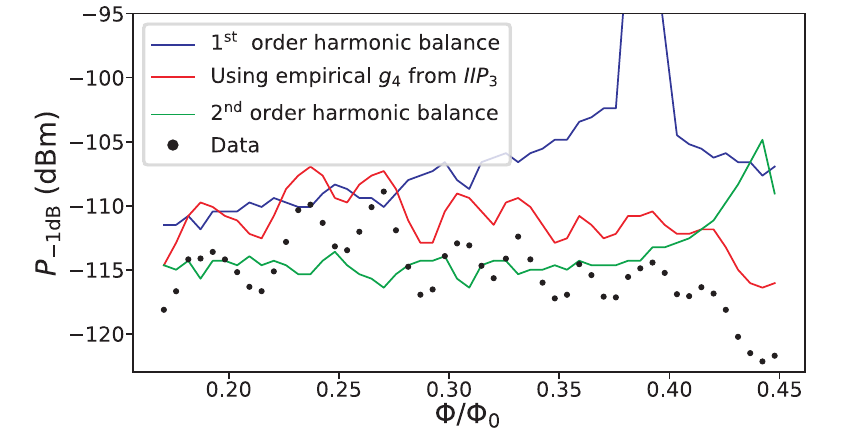}
\caption{\label{figS2_p1dB}  
Measured $P_{-1\rm{dB}}$ (black) as a function of $\Phi$ for device D biased at $20$ dB gain. Solid lines are predictions of first-order harmonic-balance theory (\ref{pump_amp})-(\ref{idler_amp}) using $g_4$ calculated from first-principles (blue) or empirically  extracted from $IIP_3$ (red). Green depicts the second-order harmonic-balance theory (\ref{pump_app_2})-(\ref{idler_app_2}) using the same parameters as blue.
}
\end{figure}
\subsection{Comparison to experimental data \label{comparison_to_data} }
In this Section, we compare the predictions of the saturation power by the harmonic-balance theory with the experimental data for device D, see Fig.~\ref{figS2_p1dB}.

Let us focus first on the blue first-order harmonic balance theory curve. To obtain it, we map the signal and pump powers onto the corresponding drive strengths $u_{s}$ and $u_{p}$, solve the system of Eqs.~(\ref{pump_amp})-(\ref{idler_amp}) numerically, and use the resulting amplitudes to calculate the gain (\ref{eq:gain semiclassics}). In this numerical procedure, we use $\omega_a$ and $\kappa$ extracted from the resonance fits at each flux point, and $g_3$ and $g_4$ calculated from first principles. As one can see from Fig.~\ref{figS2_p1dB}, the agreement with data is not satisfactory. In particular, the theory predicts a sharp peak near the Kerr-free point, which is not observed in the data.

Similarly, from Eq.~\eqref{IIP3_app} a peak near the Kerr-free point is expected in $IIP_3$, which is not observed in the data (Fig.~\ref{fig6_IMD}b). This suggests trying to instead use $g_4$ extracted from $IIP_3$ (see Fig.~\ref{figS2_g4imd}b) in our numerical calculation to predict the compression power. The result of this is the red curve on Fig.~\ref{figS2_p1dB}, which shows a much better agreement with the data at all fluxes. 

We attempt to explain the failure of the first-order harmonic balance theory near the Kerr-free point by going to higher order. We had previously included  only the main frequency components $\omega_{p}$, $\omega_{s}$ and $\omega_{i}=\omega_{p}-\omega_{s}$ in harmonic balance. The relevant frequencies to next order are $2\omega_{p}$, $2\omega_{s}$, $2\omega_{i}$, $\omega_{p}+\omega_{s}$, $\omega_{p}+\omega_{i}$, $\omega_{s}-\omega_{i}$ and $0$. The corresponding intra-resonator amplitudes of these harmonics are suppressed by a factor of $\sim\kappa/\omega_{a}$ since they fall outside the linewidth of the SPA resonator. However, increasing the population at these frequencies can cause additional Stark shifts and contribute to the reduction of the saturation power. Using the smallness of the amplitudes of these harmonics, the new extended system of harmonic balance equations can be partially solved and reduced to three equations similar to Eqs.~(\ref{pump_amp})-(\ref{idler_amp}) as shown below:
\begin{widetext}
\begin{align}
\bigg(\omega_{p}-\omega_{a}+i\frac{2}{3}\kappa-g_{4}\big[\frac{32}{9}|\alpha_{p}|^{2}+16|\alpha_{s}|^{2}+16|\alpha_{i}|^{2}\big]+\frac{g_{3}^{2}}{\omega_{c}}\big[\frac{928}{45}|\alpha_{p}|^{2}+42|\alpha_{s}|^{2}+42|\alpha_{i}|^{2}\big]\bigg)\alpha_{p}&=u_{p}+6g_{3}\alpha_{i}\alpha_{s}\label{pump_app_2}, \\
\bigg(\omega_{s}-\omega_{a}+i\frac{\kappa}{2}-g_{4}\big[\frac{32}{3}|\alpha_{p}|^{2}+12|\alpha_{s}|^{2}+12|\alpha_{i}|^{2}\big]+4\frac{g_{3}^{2}}{\omega_{c}}\big[7|\alpha_{p}|^{2}+15|\alpha_{s}|^{2}+36|\alpha_{i}|^{2}\big]\bigg)\alpha_{s}&=u_{s}+4g_{3}\alpha_{p}\alpha_{i}^{*}\label{signal_app_2}, \\
\bigg(\omega_{i}-\omega_{a}+i\frac{\kappa}{2}-g_{4}\big[\frac{32}{3}|\alpha_{p}|^{2}+12|\alpha_{i}|^{2}+12|\alpha_{s}|^{2}\big]+4\frac{g_{3}^{2}}{\omega_{c}}\big[7|\alpha_{p}|^{2}+36|\alpha_{s}|^{2}+15|\alpha_{i}|^{2}\big]\bigg)\alpha_{i}&=u_{i}+4g_{3}\alpha_{p}\alpha_{s}^{*}\label{idler_app_2}.
\end{align}
\end{widetext}

Using this system of equations with the $g_3$ and $g_4$ calculated from first principles, we are able to reconstruct the green line on Fig.~\ref{figS2_p1dB}. We can see that the agreement is significantly improved compared to the first-order harmonic balance theory (blue), except in the small region close to half flux. This second-order harmonic balance theory is compared to data across multiple devices in Fig.~\ref{fig5_P1dB}b. The remaining discrepancy (tails in Fig.~\ref{fig5_P1dB}b) can possibly be explained by going to even higher order in harmonic balance, or by including higher-order terms in the expansion of the Hamiltonian \eqref{ham_appendix}. 

Finally, we find that using the stiff-pump approximation for Stark-shift-limited saturation yields results similar to the full numerical solution. This further confirms that the effect of pump depletion is negligible and Kerr-induced Stark shifts are the primary mechanism responsible for amplifier saturation.

\bibliographystyle{apsrev_longbib} 
\bibliography{SPAexp} 

\end{document}